\documentclass[aps, twocolumn, showpacs, preprintnumbers, superscriptaddress, amsmath, amssymb, prb]{revtex4}
\usepackage{graphicx}
\usepackage{dcolumn}
\usepackage{bm}
\usepackage{color}
\usepackage{bbm}%

\makeatletter
\newsavebox\myboxA
\newsavebox\myboxB
\newlength\mylenA

\newcommand*\xoverline[2][0.75]{%
    \sbox{\myboxA}{$\m@th#2$}%
    \setbox\myboxB\null
    \ht\myboxB=\ht\myboxA%
    \dp\myboxB=\dp\myboxA%
    \wd\myboxB=#1\wd\myboxA
    \sbox\myboxB{$\m@th\overline{\copy\myboxB}$}
    \setlength\mylenA{\the\wd\myboxA}
    \addtolength\mylenA{-\the\wd\myboxB}%
    \ifdim\wd\myboxB<\wd\myboxA%
       \rlap{\hskip 0.5\mylenA\usebox\myboxB}{\usebox\myboxA}%
    \else
        \hskip -0.5\mylenA\rlap{\usebox\myboxA}{\hskip 0.5\mylenA\usebox\myboxB}%
    \fi}

\begin{document}

\title{Projective Truncation Approximation for Equations of Motion of Two-Time Green's Functions}

\author{Peng Fan}
\affiliation{Department of Physics, Renmin University of China, 100872 Beijing, China}
\author{Ke Yang}
\affiliation{Department of Physics, Renmin University of China, 100872 Beijing, China}
\author{Kou-Han Ma}
\affiliation{Department of Physics, Renmin University of China, 100872 Beijing, China}
\author{Ning-Hua Tong}
\email{nhtong@ruc.edu.cn}
\affiliation{Department of Physics, Renmin University of China, 100872 Beijing, China}
\date{\today}

\begin{abstract}
In the equation of motion approach to the two-time Green's functions, conventional Tyablikov-type truncation of the chain of equations is rather arbitrary and apt to violate the analytical structure of Green's functions. Here, we propose a practical way to truncate the equations of motion using operator projection. The partial projection approximation is introduced to evaluate the Liouville matrix. It guarantees the causality of Green's functions, fulfills the time translation invariance and the particle-hole symmetry, and is easy to implement in a computer. To benchmark this method, we study the Anderson impurity model using the operator basis at the level of Lacroix approximation. Improvement over conventional Lacroix approximation is observed. The distribution of Kondo screening in the energy space is studied using this method.
\end{abstract}

\pacs{24.10.Cn, 71.20.Be, 71.10.Fd}


\maketitle

\section{Introduction}

The Green's function (GF) method is widely used in the study of quantum many-body physics. 
Among the many different kinds of GFs, the two-time GF contains two time variables, or, in the equilibrium state, one frequency variable. The equation of motion (EOM) approach to the two-time GF has a long
history, dating back to the late 1950s.~\cite{Martin1,Bogolyubov1,Tyablikov1,Zubarev1} It is based on the Heisenberg EOM of operators. Usually, starting from a given GF, the repeated application of EOM will generate a chain of successively higher order GFs.~\cite{Hubbard1,Suhl1} At certain order this chain of GFs needs to be truncated to form a closed set of algebraic equations. The Tyablikov-type decoupling truncation~\cite{Tyablikov1} has certain arbitrariness due to the lack of guiding principles. The frequently encountered problems due to inappropriate truncation include: violation of causality in GFs, {\it i.e.}, imaginary poles in $G(z)$; violation of symmetry of Hamiltonian; insufficient or over-complete equations for the operator averages; and lack of systematics in the accuracy of results. 

One of the formal solutions to the above problems is the truncation of EOM using the idea of operator projection proposed by Mori,~\cite{Mori1,Mori2} Zwanzig,~\cite{Zwanzig1} and Tserkovnikov,~\cite{Tserkovnikov1,Tserkovnikov2,Ochoa1} {\it et al.}. In these essentially equivalent projective truncation approaches, the operators are arranged into hierarchies according to their orthogonality. Equivalently, GFs form hierarchies according to the level of correlations that they contain. At each order, the GF is expressed in a Dyson-type equation, with the "self-energy" containing higher order GFs whose operators and EOMs are defined in the space orthogonal to the subspace of lower order operators. One could carry out the calculation order by order, introducing more and more correlations and improving the results systematically.
Approaches based on this projection idea have been widely used in the study of correlated electron systems, including the two-pole approximation,\cite{Roth1} composite fermion method,\cite{Avella1,Avella2} self-consistent projection operator approach,\cite{Fulde1} operator projection method,\cite{Imada1,Imada2} and irreducible Green's function method,\cite{Kuzemsky1} {\it etc.}. The idea of projection also finds applications in nuclear model study,~\cite{Rowe1} quantum chemistry calculations,~\cite{Fulde2} and non-equilibrium quantum transport studies.\cite{Ochoa1} It has also been widely used in the study of classical fluid systems.\cite{Goetze1,Bosse1,Bosse2} 

The projective truncation theories discussed above also have difficulties. The analytical complexity of these approaches increases rapidly with the truncation order. Another difficulty is that usually the averages (or correlation functions) appearing in the projection coefficients need to be calculated self-consistently from the corresponding GFs via the spectral theorem. At higher orders, the number of such averages is so large that an analytical treatment becomes awkward, if not impossible. These difficulties confine the projection approaches to lowest several orders and strongly limit their applicability. 

In this work, we propose a systematic and practical way to implement the projection truncation for EOM of GFs. We introduce the partial projection approximation for the Liouville matrix ${\bf L}$ and reduce the calculation of ${\bf L}$ to two simpler matrices, the inner product matrix ${\bf I}$ and the natural closure matrix ${\bf M}$. By this simplification, we significantly reduce the analytical complexity and make possible the self-consistent calculation of a large number of averages. The calculation can be implemented on a modern computer and the basis can be enlarged to achieve higher accuracy. At the same time, the merit of the projection theory is fully inherited by the present scheme, including the causality of GF and systematics in the results. We use the Anderson impurity model (AIM) to demonstrate our method. Taking the numerical renormalization group (NRG)\cite{Wilson1} results as reference, we show that our results on the Lacroix operator basis are improved over the conventional Lacroix approximation.

The rest part of this paper is organized as follows. In Section II, we present the formalism of the partial projection approximation. In Section III, we apply this method to AIM and summarize the formula. The numerical results are compared with conventional Lacroix approximation and NRG in Section IV. In Section V, we discuss several issues about the proposed method and summarize this paper.

\section{Projective Truncation of EOM}

In this section, to set the framework, we first reformulate the formal projection theory for GFs in section II.A. This part is essentially equivalent to the theory of Roth.~\cite{Roth1} Our new theory is presented in section II.B, where the partial projection approximation is introduced as a practical way to carry out the projective truncation calculation.

\subsection{Formal Projective Truncation of EOM}

For a given Hamiltonian, we choose $n$ linearly independent operators to form a basis set. In the form of a column vector, these operators are organized as $\vec{A} = \left\{ A_{1}, A_{2}, ..., A_{n} \right\}^{T}$. The basis set should be chosen in such a way that the most important excitations for describing the physical properties of the system are included. When $n$ tends to the full dimension of the operator space, the results become exact because no truncation is done.

The retarded GF matrix defined by these basis operators reads
\begin{equation}      \label{1}
 {\bf G} \left( \vec{A}(t) | \vec{A}^{\dag} (t^{\prime}) \right) = -\frac{i}{\hbar} \theta(t-t^{\prime}) \left\langle \left\{ \vec{A}(t),\vec{A}^{\dag}(t^{\prime})  \right\}\right\rangle
\end{equation}
where $\theta(t-t^{\prime})$ is the Heaviside step function and $\vec{A}(t)$ is the vector of basis operators in Heisenberg picture. In this paper we only study the Fermion-type GF and the curly bracket in the above equation denotes the anti-commutator. Below we take the natural unit and drop out $\hbar$.

The equation of motion for the above GF matrix in the frequency domain reads
\begin{eqnarray}      \label{2&3}
\omega G\left( \vec{A} | \vec{A}^{\dag} \right)_{\omega} &=& \langle \{ \vec{A}, \vec{A}^{\dag} \} \rangle + G\left( [ \vec{A}, H ] |\vec{A}^{\dag} \right)_{\omega} ,   \\
\omega G\left( \vec{A}|\vec{A}^{\dag} \right)_{\omega} &=& \langle \{ \vec{A},\vec{A}^{\dag} \} \rangle - G\left( \vec{A}|[\vec{A}^{\dag}, H ] \right)_{\omega}.   
\end{eqnarray}
Usually, the commutator $[\vec{A},\hat{H}]$ ( or $[\vec{A}^{\dag},\hat{H}]$ ) contains higher order operators outside the basis. Correspondingly, new GFs describing higher order correlations are generated. Repeatedly employing EOMs to the new GFs will generate a chain of GFs until the full operator space is generated by the commutators. To solve the GFs approximately, the chain must be truncated to get a closed set of equations of GFs at a prescribed order. A notorious problem in the Tyablikov-type decoupling truncation is its arbitrariness and the subsequent violation of physical requirements, such as causality and symmetries. As an alternative, the projective truncation theory has been proposed to overcome these problems.~\cite{Roth1,Tserkovnikov1} Below we reformulate this theory in a form suitable for our subsequent treatment.

We denote the commutator as
\begin{equation}      \label{4}
[ A_{i}, H ] = \sum_{j} { \bf M}_{ji} A_{j} + B_{i}.
\end{equation}
The first term on the right-hand side contains the basis operators that naturally appear in the commutator, {\it i.e.}, natural closure part of the commutator. $B_{i}$ is the newly generated operator. We require that when written into normal product of single particle creation and annihilation operators, each additive component of $B_i$ is different from the basis operators $\{ A_k \}$. $B_i$ determined in this way is unique and non-orthogonal to the basis set $\{ A_k \}$. When symmetry considerations are used to fix $B_{i}$, some $A_k$ may be mixed in $B_i$, as shown in the treatment of particle-hole symmetry of AIM in section III. 

In order to do the operator projection, we define the inner product of two arbitrary operators $A$ and $B$ as
\begin{equation}      \label{5}
   (A|B) \equiv \langle \{ A^{\dagger}, B \} \rangle,
\end{equation}
where the average is defined as $\langle  \hat{O}\rangle = Tr(\rho\hat{O})$. In this work, $\rho$ is chosen to be the density operator of the equilibrium state of $H$ at temperature $T$, $\rho = e^{-\beta H} / Tr(e^{-\beta H} )$. There are other definitions of inner product in the literature as well.~\cite{Mori1,Mori2,Rowe1} For any operators $A$, $B$, and $C$, and coefficients $\alpha$ and $\beta$, the above definition fulfills the requirement
\begin{eqnarray}      \label{6}
&&  \left(\alpha A + \beta B | C \right) = \alpha^{\ast} (A|C) + \beta^{\ast} (B|C);     \nonumber \\
&&  \left(C| \alpha A + \beta B \right) = \alpha (C|A) + \beta (C|B);   \nonumber \\
&&  (A|B) = (B|A)^{\ast};    \nonumber \\
&&  (A|A) \geqslant 0.
 \end{eqnarray}
In the last inequality, the equal sign applies only if $A=0$.

An important property of this inner product is that the Liouville operator $\mathcal{L}$ defined by $\mathcal{L} \hat{O} = [H, \hat{O}]$ is Hermitian, {\it i.e.}, $(\mathcal{L} A| B) = (A| \mathcal{L}B)$. This conserves the time translation invariance of the equilibrium state and guarantees the causality of GFs. Note that not all inner product fulfilling Eq.(6) has this property.
The inner product matrix ${\bf I }$ of the basis operators has element
\begin{equation}      \label{7}
{\bf I}_{ij} = (A_{i} | A_{j}).
\end{equation}
$\bf{I}$ is a positive-definite Hermitian matrix.

To truncate the EOM, we project Eq.(4) to the basis operator $A_{k}$. Using the definitions ${\bf L}_{ki} \equiv (A_{k}| [A_{i}, H])$ and ${\bf P}_{ki} \equiv (A_{k}|B_{i})$, we obtain
\begin{equation}      \label{8} 
    {\bf L} = {\bf I} {\bf M} +  {\bf P}.
\end{equation} 
${\bf L}$ is the representation matrix of $(-\mathcal{L})$ in the given basis and it is Hermitian under the inner product Eq.(5). Neglecting the orthogonal component to the basis set, we can write $B_{i} \approx \sum_{j} {\bf N}_{ji} A_{j}$. Projecting this equation to $\{ A_{k} \}$ produces ${\bf P} ={\bf I}{\bf N}$. Put it into Eq.(8) and we have ${\bf L} = {\bf I} {\bf M}_{t}$, with ${\bf M}_{t} = {\bf M} + {\bf N}$ being the effective total closure matrix.
Now the EOM Eq.(4) becomes
\begin{equation}      \label{9}
 [ \vec{A}, H ] \approx {\bf M}_{t}^{T} \vec{A},
\end{equation}
and $ {\bf M}_t = {\bf I}^{-1} {\bf L}$.

Substituting Eq.(9) into Eq.(2), we obtain a closed expression for GFs,
\begin{equation}      \label{10}
G( \vec{A}|\vec{A}^{\dagger} )_{\omega} \approx \left( \omega {\bf 1} - {\bf M}_{t}^{T} \right)^{-1} { \bf I}^{T} .
\end{equation}
One can prove that substituting Eq.(9) into the right-side EOM Eq.(3) produces equivalent result. Hence the projective truncation keeps the time translation invariance of the equilibrium state. The GFs as given above have real simple poles because ${\bf M}_{t} = {\bf I}^{-1} {\bf L}$ has real eigen values. This is an important advantage compared to previous non-projective truncation schemes, such as the Tyablikov decoupling scheme.\cite{Tyablikov1} By enlarging the basis of operators, we can include more excitations in the poles of GFs and achieve higher accuracy.

For given matrices ${\bf M}_{t}$ and ${\bf I}$, the GF matrix in Eq.(10) can be obtained either from the matrix inversion $\left( \omega {\bf 1} - {\bf M}_{t}^{T} \right)^{-1}$, as done in previous analytical studies,\cite{Roth1} or by solving a generalized eigen-value problem numerically. For the latter case, suppose ${\bf U}^{-1} {\bf M}_{t} {\bf U} = {\bf \Lambda}$ gives a diagonal matrix ${\bf \Lambda} ={\text diag}\{ \lambda_1, \lambda_2, ..., \lambda_n \}$, it is easy to show that ${\bf U}$ and ${\bf \Lambda}$ are the generalized eigen-vector and eigen-value matrices of the pair of Hermitian matrices $({\bf L}, {\bf I})$,
\begin{equation}      \label{11}
   {\bf L U} = {\bf I U \Lambda}
\end{equation}
Considering $({\bf L}, {\bf I})$ being Hermitian and ${\bf I}$ positive definite,
one has real eigen values ${\bf \Lambda}$ and the generalized orthogonormal relation ${\bf U^{\dagger}I U = 1}$. The GFs can be expressed in terms of  ${\bf U}$ and ${\bf \Lambda}$ as
\begin{equation}      \label{12}
G( \vec{A}|\vec{A}^{\dagger} )_{\omega} \approx ( {\bf IU} )^{\ast} \left( \omega {\bf 1} - {\bf \Lambda}\right)^{-1} ({\bf IU})^{T}.
\end{equation}
The corresponding spectral function reads
\begin{equation}      \label{13}
   \rho(A_{i}|A_{j}^{\dagger})_{\omega} \approx  \sum_{k} (IU)^{\ast}_{ik} (IU)_{jk} \delta(\omega - \lambda_k).
\end{equation}

The calculation of GFs is thus reduced to that of two Hermitian matrices ${\bf L}$ and ${\bf I}$. Their elements contain the averages of operators on the state defined by the density matrix $\rho$ in Eq.(5). For an approximate treatment, $\rho$ can be taken as the density operator of an approximate ground state or thermal state of $H$, on which ${\bf L}$ and ${\bf I}$ can be calculated. \cite{Suhl1,Linderberg1} For a self-contained study, it is necessary to calculate ${\bf L}$ and ${\bf I}$ self-consistently with the GFs. The averages of the kind $\langle A_{j}^{\dagger} A_{i} \rangle$ can be calculated from Eq.(13) via the spectrum theorem as
\begin{equation}      \label{14}
\langle A_{j}^{\dagger} A_{i} \rangle  =  \sum_{k} \frac{ (IU)^{\ast}_{ik} (IU)_{jk} }{e^{\beta \lambda_{k}} + 1}.
\end{equation}  

Those averages not in the form of $\langle A_{j}^{\dagger} A_{i} \rangle$ need to be calculated from the EOM of additional GFs. For the average of the type $\langle \hat{O}A_{i} \rangle$ ( $\hat{O}$ is an operator outside the basis set $\{A_{k} \}$ ), the EOM leads to  
\begin{equation}      \label{15}
 \langle \hat{O} A_{i} \rangle \approx  \sum_{k} \frac{(IU)_{ik}^{\ast} \left[ U^{T} \langle \{\vec{A}, \hat{O} \} \rangle\right]_{k} }{e^{\beta \lambda_{k}} + 1}.
\end{equation}
$\langle \hat{O}A_{i} \rangle$ can then be calculated self-consistently from ${\bf I}$, ${\bf U}$, and ${\bf \Lambda}$, provided that the 
averages $\langle \{ A_{i}, \hat{O} \} \rangle$ ($i=1,2,...,n$) are linear combinations of $\{ \langle A_{k}^{\dagger} A_{p} \rangle \}$.\cite{Roth1} If this is not the case for some averages in ${\bf L}$ and ${\bf I}$, one could finally resort to the ordinary Tyablikov-type decoupling approximation to obtain them. As far as this decoupling approximation does not break the Hermiticity of  ${\bf L}$ and ${\bf I}$, the positive-definiteness of ${\bf I}$, nor the symmetries of $H$, the resulting GFs obtained from such a calculation obey the causality and symmetries of $H$.

Although the above projective truncation approximation has significant advantages over the Tyablikov-type decoupling approximation, it is faced difficulties in practice. Often, the projection coefficients ${\bf L}$ and ${\bf I}$ contain averages that cannot be calculated in a self-contained manner. Additional (uncontrolled) approximations have to be used to calculate $\bf{L}$ and/or ${\bf I}$. As the dimension $n$ of the basis set $\{ A_{i} \}$ ($i=1,2,..., n$) increases, the number of averages increases rapidly and the self-consistent calculation or additional decoupling approximation become too complicated to do, either analytically or numerically. This constraints the dimension of basis to a very small number in actual calculations, as in the two-pole approximation~\cite{Roth1} and the composite operator approach.~\cite{Avella1,Avella2}  Below, we propose an approximation scheme to simplify the calculation of ${\bf L}$ and ${\bf I}$ and partly remove the constraint.

\subsection{ Partial Projection Approximation for ${\bf L}$ }

The calculation of ${\bf I}$ is easier than ${\bf L}$. If the order of an operator $A$ is defined as the total number of single-particle creation and annihilation operators in $A$ in the normal order, the anti-commutator $\{ A_{i}^{\dagger}, A_{j} \}$ in ${ \bf I}_{ij}$ has an order $n = n_i + n_j -2$, with $n_i$ and $n_j$ being the orders of $A_{i}^{\dagger}$ and $A_{j}$, respectively. This number is smaller than the largest order of $A_{p}^{\dagger}A_{q}$. Therefore, usually the self-consistent calculation of ${\bf I}$ is feasible. The situation is different for ${\bf L}$ because ${\bf L}_{ij}$ has the largest order $n_{i}+n_{j}+2$ due to the interaction part of $H$. Hence the calculation of ${\bf L}$ needs additional consideration. 

In this section, we propose a systematic approximation for ${\bf L}$, which can maintain the correlations as much as possible and make the computation simpler. For this purpose, we make full use of the natural closure matrix ${\bf M}$ in Eq.(4). We first classify the basis operators into two groups, $\{A_1, A_2, ..., A_n \} = \{A_1^{(1)}, A_{2}^{(1)}, ..., A_{m}^{(1)} \} \cup \{A_{m+1}^{(2)}, A_{m+2}^{(2)}, ..., A_{n}^{(2)} \}$. The superscripts $(1)$ and $(2)$ denote subset-$1$ and $2$, respectively. Subset-$1$ is composed of basis operators whose commutators with $H$ close automatically. Subset-$2$ contains the rest basis operators. That is, we have $B_{i}^{(1)} = 0$ ($i=1,2,...,m$) and $B_{j}^{(2)} \neq 0$ ($j=m+1, m+2, ..., n$). Associated with this grouping of basis, the matrices ${\bf I}$, ${\bf M}$, ${\bf P}$, ${\bf N}$, and ${\bf L}$ all become $2 \times 2$ block matrices. In particular, 
\begin{equation}      \label{16}
  \mathbf{P} = \left(
\begin{array} {cc}
    \mathbf{ 0 }   & \mathbf{P}_{12}  \\
    \mathbf{ 0 }   & \mathbf{P}_{22}    
\end{array} \right),
\end{equation}
where ${\bf P}_{12}$ and ${\bf P}_{22}$ are the projection matrices from $\vec{B}^{(2)}$ to $\vec{A}^{(1)}$ and to $\vec{A}^{(2)}$, respectively. 

Under the block form of matrix representation, ${\bf L}$ becomes
\begin{equation}      \label{17}
  \mathbf{L} = \left(
\begin{array} {cc}
    \mathbf{ (IM) }_{11}  \,\,  & \mathbf{ (IM) }_{12} + \mathbf{P}_{12}   \\
    \mathbf{ (IM) }_{21}  \,\,  & \mathbf{ (IM) }_{22} + \mathbf{P}_{22}    
\end{array} \right).
\end{equation}
The Hermiticity of ${\bf L}$ leads to an exact expression for $\mathbf{P}_{12}$ as
\begin{equation}      \label{18}
 \mathbf{P}_{12} = \left[ \mathbf{ (IM) }_{21} \right]^{\dagger} - \mathbf{ (IM) }_{12}.
\end{equation}
It is equivalent to the following Hermiticity identities of ${\bf L}$,
\begin{equation}      \label{19}
   \langle \{ A_{i}^{\dagger}, [A_{j}, H] \}\rangle = \langle \{ A_{j}^{\dagger}, [A_{i}, H] \}\rangle^{\ast}, \,\,\,\,(i,j \in [1, n]).
\end{equation}
Physically, these identities describe the time translation invariance of the equilibrium state, since they are equivalent to $\langle [ \{ A_{i}^{\dagger}, A_{j} \}, H ] \rangle = i \partial \langle \{A_{i}^{\dagger}, A_{j} \} \rangle / \partial t = 0$. As an example, in Appendix A, we summarize the non-trivial identities for AIM obtained from the basis operators used in this work.

For ${\bf P}_{22}$, the Hermiticity of ${\bf L}$ gives
\begin{equation}      \label{20}
  \mathbf{P}_{22} - [\mathbf{P}_{22}]^{\dagger} =  [ \mathbf{ (IM) }_{22}]^{\dagger} - \mathbf{ (IM) }_{22} ,
\end{equation}
which only determines the anti-symmetric part of ${\bf P}_{22}$. We use a two-step projection scheme to determine the symmetric part of ${\bf P}_{22}$. That is, we assume $B^{(2)}_{i} \approx \sum_{j} [{\bf N}_{12}]_{ji} A_{j}^{(1)}$, neglecting the components orthogonal to subspace-$1$. Projecting this expression to $ A_{k}^{(1)}$ gives ${\bf N}_{12} \approx [{\bf I}_{11} ]^{-1} {\bf P}_{12}$ and projecting to $ A_{k}^{(2)}$  produces
\begin{equation}      \label{21}
   {\bf P}_{22}  \approx  {\bf I}_{21} [ {\bf I}_{11} ]^{-1} {\bf P}_{12},
\end{equation}
with the exact ${\bf P}_{12}$ in Eq.(18). We use Eq.(21) to determine the symmetric part of ${\bf P}_{22}$. 
It guarantees that the terms generated by the Tyablikov-type decoupling truncation $B_{i}^{(2)} \approx \sum_{j} c_{ji} A_{j}^{(1)}$ are contained in our approximation.

Combining Eqs.(17), (18), (20), and (21), we obtain the approximate ${\bf L}$ as
\begin{equation}       \label{22}
\mathbf{L} \approx  \mathbf{L}^{a} = \left(
\begin{array} {cc}
    \mathbf{ (IM) }_{11}  \,\,  & [ \mathbf{ (IM) }_{21}]^{\dagger}  \\
    \mathbf{ (IM) }_{21}  \,\,  & \frac{1}{2}[ \mathbf{ L}^{a}_{22} + (\mathbf{ L}^{a}_{22})^{\dagger} ]
\end{array} \right),
\end{equation}
where
\begin{equation}      \label{23}
\mathbf{ L}^{a}_{22} = ({\bf IM})_{22} + {\bf I}_{21} [ {\bf I}_{11} ]^{-1} {\bf P}_{12}.
\end{equation}
Eqs.(22) and (23) are the key approximation used in this work. Below we call it partial projection approximation. It keeps the correlation as much as possible by employing the exact conserving identities of averages. At the same time it avoids decoupling the higher order averages in ${\bf P}$. The input of the calculation are ${\bf M}$ and ${\bf I}$ matrices, which makes the scheme less arbitrary.

Before applying this scheme to AIM, we briefly overview the relation of present method to previous operator projection theories. Mori\cite{Mori1} proposed an elegant theoretical framework for calculating time correlation functions using the projection method. The theory is exact in the sense that the higher order correlations in the subspace orthogonal to the selected basis is taken into account by the generalized self-energy. In this work, we simply neglect the generalized self-energy and expect to recover the higher order correlations by expanding the basis. Continued fraction formalism was proposed for correlation functions by Mori\cite{Mori2} and Zwanzig.\cite{Zwanzig1} Our formalism can be regarded as the first level of the continued fraction in a matrix form, omitting the rest levels. We can also adapt the present theory into a many-level continued fraction formalism by using the Lanczos basis.\cite{Julien1} Therefore, the framework of this work is equivalent to previous ones. For the calculation of ${\bf L}$, previous works for Hubbard model calculated ${\bf L}$ from other approximations\cite{Suhl1,Avella1,Avella2} or from a given exact ground state.\cite{Linderberg1} In our method, all the averages are calculated self-consistently at the price of introducing the partial projection approximation for ${\bf L}$.

\section{Application to Anderson Impurity Model: Formalism}

In this section, for demonsting purpose, we apply our projective truncation scheme to AIM and derive the formalism. AIM is not only one of the best understood quantum many-body models in condensed matter physics, it is also widely used to study various physical problems including the Kondo effect,\cite{Kondo1} quantum dot physics,\cite{Pustilnik1} impurity quantum phase transition,\cite{Vojta1} and used in the dynamical mean-field theory for correlated lattice models.\cite{Vollhardt1,Kotliar1} The Hamiltonian of the AIM that we will study has the form
\begin{eqnarray}      \label{24}
\hat{H} &=& \sum_{k\sigma}\left(\epsilon_{k\sigma}-\mu\right) c_{k\sigma}^{\dag}c_{k\sigma} + \sum_{k\sigma}V_{k\sigma}\left(c_{k\sigma}^{\dag}d_{\sigma}+d_{\sigma}^{\dag}c_{k\sigma}\right)\nonumber\\ &+& \sum_{\sigma} \left(\epsilon_{d}-\mu\right)d_{\sigma}^{\dag}d_{\sigma}+ Un_{d\uparrow}n_{d\downarrow}.
\end{eqnarray}
Here, $c_{k \sigma}^{\dag}$ ($c_{k \sigma}$) is the creation (annihilation) operator of a conduction electron at $(k,\sigma)$ state with energy $\epsilon_{k \sigma}$. $d_{\sigma}^{\dag}$ ($d_{\sigma}$) is the creation (annihilation) operator of an electron with spin $\sigma$ on the impurity orbital. $\mu$ is the chemical potential. $V_{k \sigma}$ is the hybridization strength and $U$ is the Coulomb repulsion energy on the impurity orbital.
We use the Lorentzian hybridization function with a spectral function
\begin{equation}      \label{25}
  \Delta_{\sigma}(\omega ) \equiv \sum_{k} V_{k\sigma}^{2} \delta(\omega - \epsilon_{k \sigma}) 
 = \frac{\Delta \omega_{c}^{2}}{ (\omega + \sigma \delta \omega )^{2} + \omega_c^{2}}.
\end{equation}
We set $\omega_c = 1.0$ as the unit of energy. $\delta \omega$ is the magnetic bias of the bath electrons, introduced here to mimic the ferromagnetic leads in quantum dot systems and the situation of magnetic phase in the dynamical mean-field theory.
In accordance with $\Delta_{ \bar{\sigma} }(\omega) = \Delta_{\sigma}(-\omega)$, we assume that the parameters in $H$ have the following form,
\begin{eqnarray}      \label{26}
 && \epsilon_{\bar{k} \bar{\sigma}} = - \epsilon_{k \sigma} ;   \nonumber \\
 &&  V_{\bar{k} \bar{\sigma}} = V_{k \sigma}. 
\end{eqnarray}
The particle-hole symmetry of $H$ is realized at the parameter point
\begin{eqnarray}      \label{27}
&&   \epsilon_d = - U/2,   \nonumber \\
&&  \mu = 0.          
\end{eqnarray}
We implement the partial projection approximation for the operator basis at the level of Lacroix approximation\cite{Lacroix1} (Lacroix basis). The results are compared to conventional Lacroix as well as to NRG.

\subsection{basis operators}
In the work of Lacroix,\cite{Lacroix1} the higher order GF generated by the commutator of $H$ and $n_{\bar{\sigma}} d_{\sigma}$ are kept and the Tyablikov-type decoupling truncation is done for the next order EOM. Here we take the corresponding operators to form the Lacroix basis $\left\{ A_{1}, A_{2k}, A_{3}, A_{4k}, A_{5k}, A_{6k} \right\}$ ($k=1,2,..., n_{k}$), where
\begin{eqnarray}      \label{28}
&&  A^{(1)}_{1}= d_{\sigma}, \,\,\, A^{(1)}_{2k} = c_{k\sigma}, \,\,\, A^{(1)}_{3} = n_{\bar{\sigma}}d_{\sigma},   \nonumber \\
&& A^{(2)}_{4k} = n_{\bar{\sigma}} c_{k\sigma}, \,\,\, A^{(2)}_{5k} = d_{\bar{\sigma}}^{\dagger} c_{k \bar{\sigma}} d_{\sigma}, \,\,\, A^{(2)}_{6k} = c_{k\bar{\sigma}}^{\dagger} d_{\bar{\sigma}} d_{\sigma} . \nonumber \\
&&
\end{eqnarray}
The superscripts $(1)$ and $(2)$ denote the grouping of basis operators according to the closure properties of their commutators with $H$: $B^{(1)}_{i} = 0$ and $B^{(2)}_{i} \neq 0$.  The inner product matrix ${\bf I}$ is written into a $2\times 2$ block matrix. The sub-matrices are
\begin{eqnarray}      \label{29}
{\bf I}_{11} &=&
\left(\begin{array}{ccc}
1  &  0  & \langle n_{\bar{\sigma}} \rangle    \\
0  &  \delta_{kp}  &  0    \\
\langle n_{\bar{\sigma}} \rangle  &  0  & \langle n_{\bar{\sigma}} \rangle
\end{array}
\right),
\end{eqnarray}
\begin{eqnarray}      \label{30}
{\bf I}_{12} &=& [{\bf I}_{21}]^{\dagger} =
\left(\begin{array}{ccc}
0  &   \langle d_{\bar{\sigma}}^{\dagger} c_{p \bar{\sigma}} \rangle  & \langle d_{\bar{\sigma}}^{\dagger} c_{p \bar{\sigma}} \rangle    \\
\langle n_{\bar{\sigma}} \rangle \delta_{kp}  &  0  &  0    \\
0  &  \langle d_{\bar{\sigma}}^{\dagger} c_{p \bar{\sigma}} n_{\sigma} \rangle  &  \langle d_{\bar{\sigma}}^{\dagger} c_{p \bar{\sigma}} (1-n_{\sigma}) \rangle
\end{array}
\right),      \nonumber \\
&&
\end{eqnarray}
and
\begin{eqnarray}      \label{31}
{\bf I}_{22} &=& 
\left(\begin{array}{ccc}
\langle n_{\bar{\sigma}} \rangle \delta_{kp}  &   \langle d_{\bar{\sigma}}^{\dagger} c_{p \bar{\sigma}} c_{k\sigma}^{\dagger} d_{\sigma} \rangle  & \langle d_{\bar{\sigma}}^{\dagger} c_{p \bar{\sigma}} c_{k\sigma} d_{\sigma}^{\dagger} \rangle   \\
\langle d_{\bar{\sigma}}^{\dagger} c_{k \bar{\sigma}} c_{p\sigma}^{\dagger} d_{\sigma} \rangle   &  W_{kp}  &  0    \\
\langle d_{\bar{\sigma}}^{\dagger} c_{k \bar{\sigma}} c_{p\sigma} d_{\sigma}^{\dagger} \rangle  &  0  &  U_{kp}
\end{array}
\right).   \nonumber \\
&&   
\end{eqnarray}
In ${\bf I}_{22}$, the diagonal elements $W_{kp}$ and $U_{kp}$ read
\begin{eqnarray}      \label{32}
   W_{kp} &=& \langle c_{k \bar{\sigma}}^{\dagger} c_{p\bar{\sigma}}(n_{\sigma} - n_{\bar{\sigma}} )  \rangle + \langle n_{\bar{\sigma}} ( 1 - n_{\sigma} ) \rangle \delta_{kp},     \nonumber \\
   U_{kp} &=& \langle c_{p \bar{\sigma}}^{\dagger} c_{k\bar{\sigma}}( 1 - n_{\sigma} - n_{\bar{\sigma}} )  \rangle + \langle n_{\bar{\sigma}} n_{\sigma}  \rangle \delta_{kp}.
\end{eqnarray}
Here, each element with a subscript $k$ and/or $p$ represents a matrix with $k$ and $p$ being rank and column indices, respectively. For an example, $\delta_{kp}$ represents a unity matrix of size $n_k \times n_k$.

\subsection{particle-hole symmetric formalism}

Before applying the partial projection approximation to ${\bf L}$, we need first discuss the particle-hole symmetry of the formalism. In general, neither the full projection approximation, {\it i.e.}, without additional approximation for ${\bf L}$, nor the partial projection approximation for ${\bf L}$ guarantees the particle-hole symmetry of $H$. We find that direct use of the partial projection approximation Eqs.(22) and (23) for Lacroix basis produces results that weakly violate the particle-hole symmetry. Below, we adapt the formalism of partial projection approximation to a particle-hole symmetric form, following the idea of Ref.~\onlinecite{Dorneich1}. 

The particle-hole transformation for AIM Eq.(24) is defined as
\begin{eqnarray}      \label{33}
&& d_{\sigma} \rightarrow  d^{\prime}_{\sigma}  \equiv  d_{\bar{\sigma}}^{\dagger},   \nonumber \\
&& c_{k \sigma} \rightarrow  c^{\prime}_{k \sigma} \equiv  - c_{\bar{k} \bar{\sigma}}^{\dagger}.     
\end{eqnarray}
We have $H^{\prime}=H$ at the parameter conditions Eqs.(26) and (27). In order to make the theory particle-hole symmetric and at the same time only keep the annihilation operators in the basis, we introduce the composite transformation
\begin{equation}      \label{34}
 O \rightarrow \tilde{O}  \equiv  \left( \bar{ O }^{\prime}  \right)^{\dagger} ,
\end{equation}
where the transformation $\bar{O} = O |_{k \rightarrow \bar{k}, \sigma \rightarrow \bar{\sigma}}$ applies both to the operators and to the parameters contained in $O$. For a parameter $\alpha_{k \sigma}$, we have $\tilde{ \alpha}_{k, \sigma} = (\alpha_{\bar{k}, \bar{\sigma}})^{\ast}$. For the average of an operator $\langle O \rangle$, the above transformation is defined as $\widetilde{ \langle O \rangle } = \langle \tilde{ O} \rangle$ without changing the state on which the average is evaluated. The three operations in this composite transformation commute with each other. Eq.(34) removes the additional effect of Hermitian conjugate and the inversion of momentum and spin from the particle-hole transformation Eq.(33) and it keeps the basis in the subspace of annihilation operators. It is easy to prove that $\tilde{H} = H$ at the particle-hole symmetric point. 

The subspace spanned by Lacroix basis is invariant under the transformation Eq.(34) and the basis operators change according to $\vec{ \tilde{A} }  = {\bf Q} \vec{A}$, with
\begin{eqnarray}      \label{35}
{\bf Q} &=& 
\left(\begin{array}{cccccc}
1  &  0  &  0   &   0  &   0  &   0   \\
0  &  -1  &  0   &   0  &   0  &   0   \\
1  &  0  &  -1  &   0  &   0  &   0   \\
0  &  -1  &  0   &  1  &   0  &   0   \\
0  &  0  &  0   &   0  &   1  &   0   \\
0  &  0  &  0   &   0  &   0  &   1   \\
\end{array}
\right).
\end{eqnarray}
In this matrix, the number represents a block matrix if the basis operators have bath index $k$. Because two successive composite transformations make $\vec{A}$ unchanged, Eq.(35) fulfills ${\bf Q}^{\ast}{\bf Q} = {\bf 1}$ . A particle-hole symmetric formalism should be invariant under this composite transformation. 

For general parameters, $H$ can be decomposed into $H= H_{e} + H_{o}$. $H_e = (H + H^{\prime})/2$ is even under the particle-hole transformation, $H_{e}^{\prime} = H_{e}$. $H_o = (H - H^{\prime})/2$ is odd, $H_{o}^{\prime} = - H_{o}$.
Correspondingly, we have ${\bf L} = {\bf L}_{e} + {\bf L}_{o}$ and  ${\bf M} = {\bf M}_{e} + {\bf M}_{o}$ where the subscript $e$ ($o$) denotes quantities calculated using the commutators with $H_{e}$ ($H_{o}$). The relation ${\bf L}_{\alpha} = {\bf I} {\bf M}_{\alpha} + {\bf P}_{\alpha}$ holds for both components $\alpha = e$ and $o$. Note that the definition of inner product Eq.(5) still uses the full $H$. Examining the properties of matrices ${\bf I}_{\alpha}$, ${\bf M}_{\alpha}$, ${\bf P}_{\alpha}$, and ${\bf L}_{\alpha}$ ($\alpha = e, o$) under the composite transformation, we obtain
\begin{equation}      \label{36}
  \tilde{ {\bf I} } = {\bf Q}^{\ast} {\bf I} {\bf Q}^{T},
\end{equation} 
\begin{eqnarray}      \label{37}
&&  \tilde{ {\bf M} }_{e} = - ({\bf Q}^{\ast})^{T} {\bf M}_{e} {\bf Q}^{T},  \nonumber \\ 
&&  \tilde{ {\bf P} }_{e} = - {\bf Q}^{\ast} {\bf P}_{e} {\bf Q}^{T},  \nonumber \\  
&&  \tilde{ {\bf L} }_{e} = - {\bf Q}^{\ast} {\bf L}_{e} {\bf Q}^{T},
\end{eqnarray}
and
\begin{eqnarray}      \label{38}
&&  \tilde{ {\bf M} }_{o} = ({\bf Q}^{\ast})^{T} {\bf M}_{o} {\bf Q}^{T},  \nonumber \\ 
&&  \tilde{ {\bf P} }_{o} = {\bf Q}^{\ast} {\bf P}_{o} {\bf Q}^{T},  \nonumber \\  
&&  \tilde{ {\bf L} }_{o} = {\bf Q}^{\ast} {\bf L}_{o} {\bf Q}^{T}.
\end{eqnarray}
Obviously, $\tilde{ {\bf L} }_{\alpha} = \tilde{ {\bf I} } \tilde{ {\bf M} }_{\alpha} + \tilde{ {\bf P} }_{\alpha}$ ($\alpha = e, o$). To obtain these relations, we have required that $\vec{ \tilde{B} }_{e} = - {\bf Q} \vec{B}_{e} $ and  $\vec{ \tilde{B} }_{o} = {\bf Q} \vec{B}_{o} $, which may mix some basis operators in them. lead to linear dependence of them on the basis operators. This is the essential ingredients for producing a particle-hole symmetric form for ${\bf M}_{\alpha}$ and ${\bf P}_{\alpha}$. We can prove that the particle-hole symmetry is conserved if ${\bf I}$ and ${\bf L}$ used in Eq.(10) obeys Eqs.(36) and (37). The proof is given in Appendix C. The partial projection approximation for ${\bf L}$, however, does not respect Eqs.(37) and (38). This leads to a slight breaking of particle-hole symmetry in the results. A symmetrization procedure has to be used together with the partial projection approximation.

Below we develop a particle-hole symmetric formalism for the partial projection approximation. We define an Hermitian Liouville matrix in the transformed space ${\bf L}_{s} \equiv - \tilde{ {\bf L} }_{e} + \tilde{ {\bf L} }_{o} = {\bf Q}^{\ast} {\bf  L} {\bf Q}^{T}$ and make the partial projection approximation ${\bf L}_{s} \approx \left( {\bf L}_{s} \right)^{a}$, in the same way as for ${\bf L} \approx {\bf L}^{a}$ in Eq.(22). 

Since in Eq.(22) ${\bf L}^{a}$ is expressed in terms of ${\bf I}$ and ${\bf M}$ only, we write ${\bf L}_{s} = \tilde {{\bf I} } {\bf M}_{s} + {\bf P}_{s}$ and approximate it using Eqs.(22) and (23), replacing ${\bf I}$ and ${\bf M}$ with $\tilde{ {\bf I} }$ and ${\bf M}_{s}$, respectively. Here ${\bf M}_{s} = ({\bf Q}^{\ast})^{T}{\bf M}{\bf Q}^{T}$ and ${\bf P}_{s} = {\bf Q}^{\ast} {\bf P} {\bf Q}^{T}$. The particle-hole-symmetry-conserving partial projection approximation is finally obtained by the following symmetrizing procedure
\begin{equation}      \label{39}
   {\bf L} = \frac{1}{2} \left[ {\bf L}^{a} +  {\bf Q} \left( {\bf L}_{s}\right)^{a} ({\bf Q}^{\ast})^{T}  \right].
\end{equation}
It is easy to check that this formula for ${\bf L}$ fulfills the requirement Eqs.(37) and (38) in the limits $H_{o}=0$ and $H_{e}=0$, respectively, and hence produce particle-hole symmetric GFs according to Appendix C.

\subsection{ {\bf M} matrix }

To calculate ${\bf M}$ for the Lacroix basis, we use ${\bf M} ={\bf M}_{e} + {\bf M}_{o} $. ${\bf M}_{\alpha}$ ($\alpha = e, o$) is associated with $\vec{B}_{\alpha}$ through the definition Eq.(4). The particle-hole symmetric and anti-symmetric components of $H$ are obtained as
\begin{eqnarray}      \label{40}
   H_{e} &=& \sum_{k, \sigma} \left[ \epsilon_{k \sigma} c_{k\sigma}^{\dagger} c_{k \sigma} + V_{k \sigma} (d_{\sigma}^{\dagger} c_{k \sigma} + c_{k \sigma}^{\dagger} d_{\sigma} ) \right]          \nonumber \\
  & &  - \frac{U}{2} \sum_{\sigma} n_{\sigma} + U{n}_{\uparrow}n_{\downarrow} + c,  \nonumber \\ 
  H_{o} &=& - \mu \sum_{k, \sigma} c_{k\sigma}^{\dagger} c_{k \sigma} + \left( \epsilon_d - \mu + \frac{U}{2} \right) \sum_{\sigma}  n_{\sigma} -c,   \nonumber \\
  &&
\end{eqnarray}
where $c = \epsilon_d - \mu + U/2 - \mu \sum_{k} 1$.
Using the commutators $[A_{i}, H_{\alpha}]$ ($\alpha = e, o$) in Appendix B, we obtain
\begin{eqnarray}      \label{41}
{\bf M}_{e} &=& 
\left(\begin{array}{cccccc}
-\frac{U}{2}  &  V_{p \sigma}  &   0    &   0  &   -\frac{V_{p \bar{\sigma}}}{2}  &   \frac{V_{p \bar{\sigma}}}{2}   \\
V_{k \sigma}  &  \epsilon_{k\sigma} \delta_{kp}  &  0   &   0  &   0  &   0   \\
U  &  0  &  \frac{U}{2}  &   V_{p \sigma}  &   V_{p \bar{\sigma}}  &   -V_{p \bar{\sigma}}   \\
0  &  0  &  V_{k \sigma}   &  \epsilon_{k \sigma}\delta_{kp}  &   0  &   0   \\
0  &  0  &  V_{k \bar{\sigma}}   &   0  &  \epsilon_{k \bar{\sigma}}\delta_{kp}  &   0   \\
0  &  0  &  -V_{k \bar{\sigma}}  &   0  &   0  &  -\epsilon_{k \bar{\sigma}}\delta_{kp}  \\
\end{array}
\right),   \nonumber \\
&&
\end{eqnarray}
and

\begin{eqnarray}      \label{42}
{\bf M}_{o} &=& 
\left(\begin{array}{cccccc}
\bar{\epsilon}_d  &  0  &   0    &   0  &   0  &  0   \\
0  &  -\mu \delta_{kp}  &  0   &   0  &   0  &   0   \\
0  &  0  &  \bar{\epsilon_d}  &   0  &   0  &   0  \\
0  &  0  &  0  &  -\mu \delta_{kp}  &   0  &   0   \\
0  &  0  &  0  &   0  &  -\mu \delta_{kp}  &   0   \\
0  &  0  &  0  &   0  &   0  &  (2\bar{\epsilon}_{d}+\mu) \delta_{kp}  \\
\end{array}
\right).   \nonumber \\
&&
\end{eqnarray}
Here, $\bar{\epsilon}_d= \epsilon_d - \mu + U/2$. We have used the parameter condition Eq.(26) to simplify the above equations.
The new operators are obtained as $\vec{B}_{o} =0$ and 
\begin{eqnarray}      \label{43}
&&  (B_{e})_{1} =(B_{e})_{2k}=(B_{e})_{3} =0 ,   \nonumber \\ 
&&  (B_{e})_{4k} = \sum_{p} V_{p \bar{\sigma}} \left( d_{\bar{\sigma}}^{\dagger} c_{p \bar{\sigma}} c_{k \sigma} - c_{p \bar{\sigma}}^{\dagger}  d_{\bar{\sigma}} c_{k \sigma}  \right) ,   \nonumber \\
&&  (B_{e})_{5k} = \sum_{p} \left[ V_{p \sigma} d_{\bar{\sigma}}^{\dagger} c_{k \bar{\sigma}} c_{p \sigma} - V_{p \bar{\sigma}}\left( c_{p \bar{\sigma}}^{\dagger} c_{k \bar{\sigma}} -\frac{1}{2}\delta_{kp} \right) d_{\sigma} \right] ,   \nonumber \\
&&  (B_{e})_{6k} = \sum_{p} \left[ V_{p \sigma} c_{k \bar{\sigma}}^{\dagger} d_{\bar{\sigma}} c_{p \sigma} + V_{p \bar{\sigma}}\left( c_{k \bar{\sigma}}^{\dagger} c_{p \bar{\sigma}} -\frac{1}{2}\delta_{kp} \right) d_{\sigma} \right].  \nonumber \\
&& 
\end{eqnarray}
Note that in identifying $\vec{B}_e$ and $\vec{B}_o$ from the expressions of commutators, we used their symmetry requirements $\vec{ \tilde{B} }_{e} = - {\bf Q} \vec{B}_{e} $ and  $\vec{ \tilde{B} }_{o} = {\bf Q} \vec{B}_{o} $.

\subsection{ self-consistent calculation of ${\bf I}$ }

The averages appearing in the inner product matrix Eqs.(29)-(32) include 
\begin{eqnarray}      \label{44}
&& \langle n_{\bar{\sigma}} \rangle  = \langle A_{1 \bar{\sigma}}^{\dagger}A_{1 \bar{\sigma}} \rangle,   \nonumber \\
&& \langle n_{\bar{\sigma}} n_{\sigma} \rangle  = \langle A_{1 \sigma}^{\dagger} A_{3 \sigma} \rangle,   \nonumber \\
&& \langle d_{\bar{\sigma}}^{\dagger} c_{p\bar{\sigma}} \rangle  = \langle A_{1 \bar{\sigma}}^{\dagger} A_{2p \bar{\sigma}} \rangle,   \nonumber \\ 
&& \langle d_{\bar{\sigma}}^{\dagger} c_{p\bar{\sigma}}n_{\sigma} \rangle  = \langle A_{3 \bar{\sigma}}^{\dagger} A_{2p \bar{\sigma}} \rangle,   \nonumber \\ 
&& \langle d_{\bar{\sigma}}^{\dagger} c_{p\bar{\sigma}} c_{k \sigma}^{\dagger} d_{\sigma} \rangle  = \langle A_{2k \sigma}^{\dagger} A_{5p \sigma} \rangle,   \nonumber \\ 
&& \langle d_{\bar{\sigma}}^{\dagger} c_{p\bar{\sigma}} c_{k \sigma} d_{\sigma}^{\dagger} \rangle  = - \langle A_{2k \sigma}^{\dagger} A_{6p \sigma}^{\ast} \rangle,   \nonumber \\ 
&& \langle c_{k \bar{\sigma}}^{\dagger} c_{p \bar{\sigma}}  n_{\sigma} \rangle = \langle A_{2k \bar{\sigma}}^{\dagger} A_{4p \bar{\sigma}} \rangle.
\end{eqnarray} 
They are written in the form $\langle A_{j}^{\dagger} A_{i}\rangle$ and can be calculated self-consistently from Eq.(14). The only average not in this form is $\langle c_{k \bar{\sigma}}^{\dagger} c_{p \bar{\sigma}}  n_{\bar{\sigma}} \rangle$ in Eq.(32), which has to be calculated from Eq.(15). For this purpose, we write $\langle c_{k \bar{\sigma}}^{\dagger} c_{p \bar{\sigma}}  n_{\bar{\sigma}} \rangle = \langle \hat{O}_{k \bar{\sigma}} A_{2p \bar{\sigma}}\rangle$ with $\hat{O}_{k \bar{\sigma}} =  c_{k \bar{\sigma}}^{\dagger} n_{\bar{\sigma}}$. Since the averages $\langle \{ \vec{A}_{\bar{\sigma}}, \hat{O}_{k \bar{\sigma}} \} \rangle$ can be written in the form  $\langle A_{j}^{\dagger} A_{i}\rangle$, $\langle c_{k \bar{\sigma}}^{\dagger} c_{p \bar{\sigma}}  n_{\bar{\sigma}} \rangle$ can also be calculated self-consistently. For details, we have
\begin{eqnarray}      \label{45}
\langle \{A_{1 \bar{\sigma}}, \hat{O}_{k \bar{\sigma}} \} \rangle &=& - \langle d_{\bar{\sigma}}^{\dagger} c_{k \bar{\sigma}}\rangle = - \langle A_{1 \bar{\sigma} }^{\dagger} A_{2k \bar{\sigma} } \rangle  ,  \nonumber \\
\langle \{A_{2p\bar{\sigma}}, \hat{O}_{k \bar{\sigma}} \} \rangle &=& \langle n_{\bar{\sigma}} \rangle \delta_{kp} =  \langle A_{1 \bar{\sigma} }^{\dagger} A_{1 \bar{\sigma} } \rangle \delta_{kp}  , \nonumber \\
\langle \{A_{3\bar{\sigma}}, \hat{O}_{k \bar{\sigma}} \} \rangle &=& - \langle n_{\sigma} d_{\bar{\sigma}}^{\dagger} c_{k \bar{\sigma}}\rangle = - \langle A_{3 \bar{\sigma} }^{\dagger} A_{2k \bar{\sigma} } \rangle ,  \nonumber \\
\langle \{A_{4p\bar{\sigma}}, \hat{O}_{k \bar{\sigma}} \} \rangle &=& \langle n_{\bar{\sigma}} n_{\sigma} \rangle \delta_{kp} = \langle A_{1 \bar{\sigma} }^{\dagger} A_{3 \bar{\sigma} } \rangle \delta_{kp} ,  \nonumber \\
\langle \{A_{5p\bar{\sigma}}, \hat{O}_{k \bar{\sigma}} \} \rangle &=& \langle d_{\sigma}^{\dagger} c_{p \sigma} d_{\bar{\sigma}} c_{k \bar{\sigma}}^{\dagger}\rangle = - \langle A_{2k \bar{\sigma} }^{\dagger} A_{5p \bar{\sigma} } \rangle ,  \nonumber \\
\langle \{A_{6p\bar{\sigma}}, \hat{O}_{k \bar{\sigma}} \} \rangle &=& \langle c_{p \sigma}^{\dagger} d_{ \sigma} d_{\bar{\sigma}} c_{k \bar{\sigma}}^{\dagger} \rangle = - \langle A_{2k \bar{\sigma} }^{\dagger} A_{6p \bar{\sigma} } \rangle .  \nonumber \\
&& 
\end{eqnarray}

Therefore, all the involved averages are calculated self-consistently.
Note that an operator could have inequivalent ways to be split into the form $A_{j}^{\dagger}A_{i}$. Different splitting may lead to different converged results of the averages. In our calculation, we found that different ways of splitting give slight deviations on the order of $10^{-4}$ and is not a severe problem. In case that the Hermiticity of ${\bf I}$ is slightly broken by the arbitrariness in the splitting, we simply symmetrize ${\bf I}$ to remove this effect.

\subsection{bath discretization}

For the numerical treatment for Lacroix basis, we need to discretize the bath degrees of freedom. To compromise the energy resolution and the discretization error, we use similar discretization formula as used in NRG\cite{Wilson1,Bulla1} but on a power-law energy mesh. We split the full energy window $[-D, D]$ into $n_{k} = 2N+1$ intervals that distribute symmetrically, with $N$ intervals on the positive energy side, another $N$ on the negative side, and an interval covering zero energy in the middle. Here $D$ is the cutoff energy. Once the energy mesh $\{ \omega_{i} \}$ is fixed, following NRG formalism,\cite{Bulla1} the continuous bath degrees of freedom in the $i$-th energy interval $[\omega_{i+1}, \omega_i]$ ($i=1,2,...,n_{k}$) are represented by a single bath site with parameters 
\begin{eqnarray}      \label{46}
 &&  V_{i \sigma} = \left[ \int_{\omega_{i+1}}^{\omega_i} \Delta_{\sigma}(\epsilon) d \epsilon \right]^{1/2},   \nonumber \\
&&   \epsilon_{i \sigma} = \frac{  \int_{\omega_{i+1}}^{\omega_i} \epsilon \Delta_{\sigma}(\epsilon) d \epsilon }{\int_{\omega_{i+1}}^{\omega_i} \Delta_{\sigma}(\epsilon) d \epsilon } . 
\end{eqnarray}

There are different ways of choosing the mesh, such as the logarithmic discretization used in NRG. Here, for the positive energy side, we set the length of the $i$-th interval $[\omega_{i+1}, \omega_i]$ ($i=1, 2, ..., N$)
\begin{equation}      \label{47}
\Delta_{i}= \omega_{i}-\omega_{i+1} = c / i^{s}, \,\,\,\,\,\, (i=1, 2, ..., N).
\end{equation}
Here $s$ is the power index. For the central interval $[\omega_{N+2}, \omega_{N+1}] = [-\omega_{N+1}, \omega_{N+1}]$, we set its length $\Delta_{N+1} = c/ (N+1)^{s}$. The constant $c$ is therefore fixed by
\begin{equation}      \label{48}
  c \left( \sum_{i=1}^{N} \Delta_{i} + \frac{1}{2} \Delta_{N+1} \right) = D.
\end{equation}
The left boundaries of the intervals on the positive side are thus obtained recursively by $\omega_1 = D$ and 
\begin{equation}      \label{49}
   \omega_{i+1} = \omega_i - \Delta_{i}, \,\,\,\,\,\, (i=1, 2, ..., N).
\end{equation}
We have $\omega_{N+2} = -\omega_{N+1}$ for the left boundary of the central interval. For the negative energy side, the left boundary of intervals are given by
\begin{equation}      \label{50}
   \omega_{N+2+i} = - \omega_{N+1-i}, \,\,\,\,\,\, (i=1, 2, ..., N)
\end{equation}
By tuning the index $s$, we can obtain an optimal distribution of bath sites in energy space such that both the low energy Kondo peak and the high energy Hubbard peaks have a satisfactory resolution. Empirically, we find the optimal value $s=0.0$, {\it i.e.}, a uniform discretization works best for most situations. When it is necessary for a high resolution of the Kondo peak, we use $s = 0.3 \sim 2.0$.

\section{Application to Anderson Impurity Model: Results}

Using the formalism developed in previous sections, we obtain numerical results for the Lacroix basis. Below we compare the results from the partial projection approximation on Lacroix basis (pLacroix) with those from conventional Lacroix approximation (cLacroix) and NRG. 

NRG results are obtained from the full density matrix NRG algorithm\cite{Weichselbaum1,Peters1,Fang2} with logarithmic discretization parameter $\Lambda=2.0$ and the number of kept states $M_s = 350 \sim 380$. For the local density of states (LDOS), we use the self-energy trick\cite{Bulla2} and average the results on $N_{z}=8$ interleaved discretizations.\cite{Yoshida1} Though not extrapolated to the exact limit $\Lambda=1$ and $M_{s}=\infty$,\cite{Li1} we have checked that the uncertainties in NRG results are much smaller than the difference between NRG and all approximate results.

In the calculation below, we fix the chemical potential $\mu=0.0$ and the hybridization strength $\Delta=0.1$. For the matrix calculation for Lacroix basis, we choose $D=5.0$ and $n_{k}=401$. We use the power law discretization index $s=0.0$ and broaden the $\delta$-peaks in LDOS with $\eta=0.01 \sim 0.02$, unless stated otherwise.

\subsection{comparisons among pLacroix, cLacroix and NRG}

In this subsection, we compare the results of pLacroix, cLacroix, and NRG. We study $\langle n_{\sigma} \rangle$ and $\langle n_{\uparrow} n_{\downarrow} \rangle$ as functions of $\epsilon_d$, $\delta \omega$, and $T$. They describe the magnetic and the charge response properties of AIM as functions of external parameters. We also study the evolution of LDOS with parameters $U$, $\epsilon_d$, and $T$. 
\begin{figure}[t!]
\vspace{-1.0cm}
\begin{center}
\includegraphics[width=5.0in, height=3.8in, angle=0]{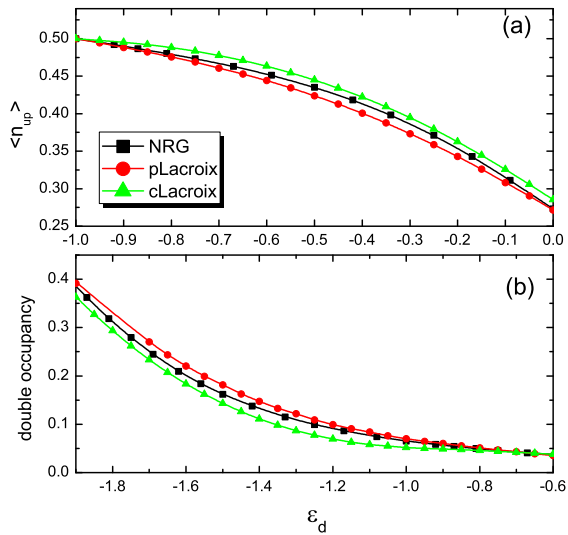}
\vspace*{-2.5cm}
\end{center}
\caption{(a) $\langle n_{\uparrow} \rangle$ and (b) $\langle n_{\uparrow} n_{\downarrow}\rangle$ as functions of impurity energy level $\epsilon_d$. Other parameters are $U=2.0$, $T=0.1$, $\delta \omega = 0.0$, and $\Delta =0.1$. }   \label{Fig1}
\end{figure}
Fig.1 shows $\langle n_{\uparrow} \rangle$ and double occupancy $\langle n_{\uparrow} n_{\downarrow} \rangle$ as functions of $\epsilon_d$, Compared to pLacroix, the result of cLacroix deviates from NRG more severely in the small and the large $\epsilon_d$ regimes. Note that pLacroix underestimates $\langle n_{\uparrow}\rangle$ and overestimates $\langle n_{\uparrow} n_{\downarrow} \rangle$, being contrary to cLacroix.

\begin{figure}[t!]
\vspace{-1.0cm}
\begin{center}
\includegraphics[width=5.0in, height=3.8in, angle=0]{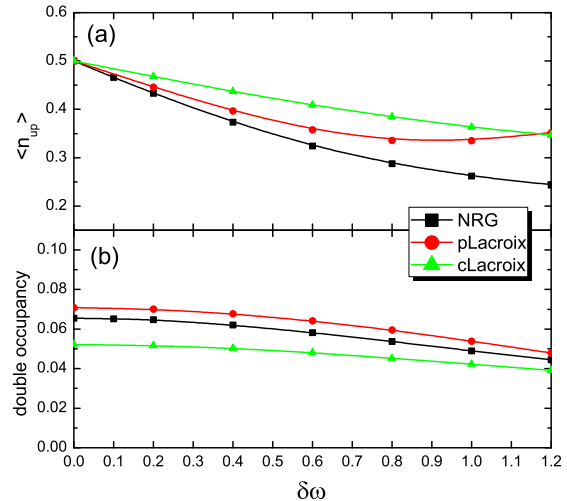}
\vspace*{-3.0cm}
\end{center}
\caption{(a) $\langle n_{\uparrow} \rangle$ and (b) $\langle n_{\uparrow} n_{\downarrow}\rangle$ as functions of magnetic bias $\delta \omega$ of the bath. Other parameters are $U=2.0$, $\epsilon_d = -U/2$, $T=0.1$, and $\Delta =0.1$. }   \label{Fig2}
\end{figure}

\begin{figure}[t!]
\vspace{-1.0cm}
\begin{center}
\includegraphics[width=5.0in, height=3.8in, angle=0]{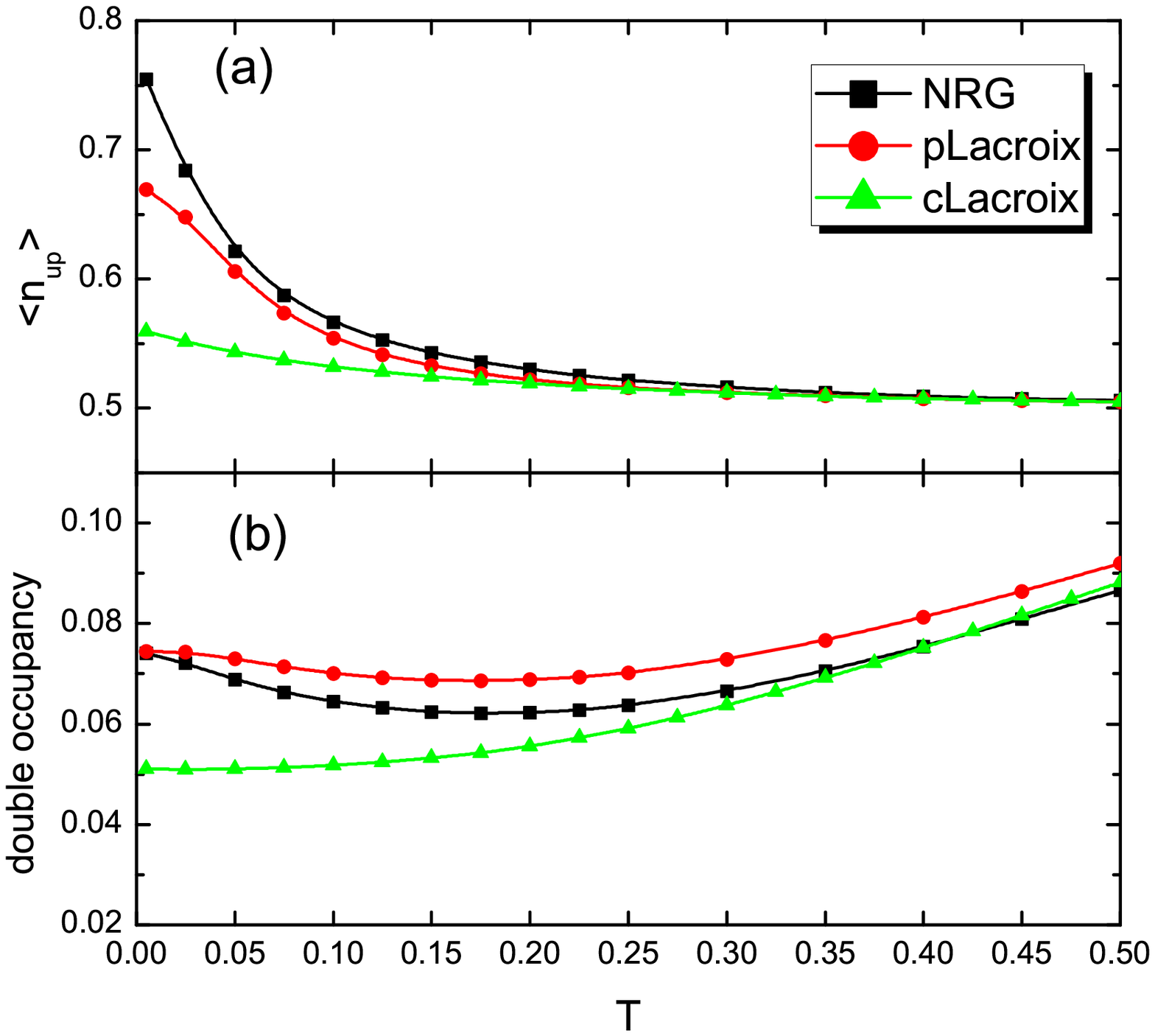}
\vspace*{-3.0cm}
\end{center}
\caption{(a) $\langle n_{\uparrow} \rangle$ and (b) $\langle n_{\uparrow} n_{\downarrow}\rangle$ as functions of $T$. Other parameters are $U=2.0$, $\epsilon_d = -U/2$, $\delta \omega = -0.2$, and $\Delta =0.1$.  }   \label{Fig3}
\end{figure}

In Fig.2, the same quantities are plotted as functions of $\delta \omega$ at $U=2.0$, $\epsilon_d = -U/2$, $T=0.1$, and $\Delta =0.1$. The particle-hole symmetry guarantees $\langle n_{\uparrow}\rangle + \langle n_{\downarrow} \rangle=1$. The curve $\langle n_{\uparrow} \rangle$-$\delta\omega$ describes the response of the impurity spin to the magnetic field on the bath, which is important for studying the magnetic phase of Hubbard model through DMFT. Since $\delta \omega > 0$ corresponds to a decrease of $\epsilon_{k\uparrow}$, it leads to positive bath polarization $\langle n_{k z} \rangle > 0$. The anti-ferromagnetic Kondo coupling between the impurity and the bath spins predicts $\langle n_{d z} \rangle < 0$, {\it i.e.}, $\langle n_{\uparrow} \rangle < 0.5$. As shown in Fig.2(a), pLacroix, cLacroix, and NRG produce $\langle n_{\uparrow} \rangle < 0.5$, consistent with the prediction. pLacroix result agree better with NRG in the regime $\delta  \omega < 1.2$. In the larger $\delta \omega$ regime, both pLacroix and cLacroix give smaller impurity polarization as compared to NRG, with pLacroix result being less accurate. In Fig.2(b), all the methods produce weakly $\delta \omega$-dependent double occupancies and pLacroix is better than cLacroix.

The temperature dependence of the same quantities are shown in Fig.3 for $U=2.0$, $\epsilon_d=-U/2$, $\delta\omega = -0.2$, and $\Delta=0.1$. In Fig.3(a), the impurity spin polarization increases as temperature is lowered for both pLacroix and cLacroix, with the former being closer to NRG result. The double occupancies shown in Fig.3(b) have similar trend. It is noted that pLacroix produces the slight increase of double occupancy as $T$ decreases below $0.15$. This upturn of double occupancy, also seen in NRG result, is associated with the screening of local moment and forming of the Fermi liquid state when temperature decreases below the Kondo temperature.

\begin{figure}[t!]
\vspace{-1.0cm}
\begin{center}
\includegraphics[width=5.2in, height=3.8in, angle=0]{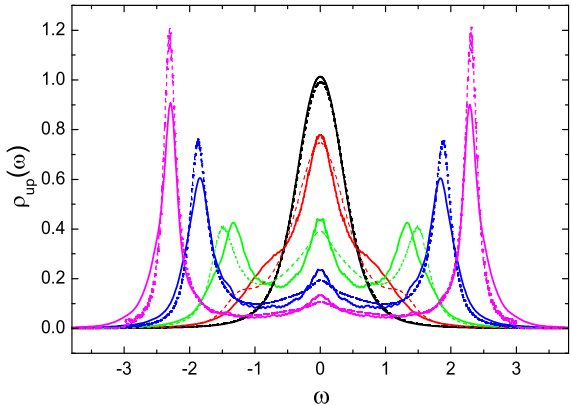}
\vspace*{-3.5cm}
\end{center}
\caption{Comparison of LDOS from pLacroix (dashed lines) and NRG (solid lines). From top to bottom at $\omega=0$, $U=0.0$, $1.0$, $2.0$, $3.0$, and $4.0$, respectively. Other parameters are $T=0.1$, $\epsilon_d=-U/2$, $\delta \omega = 0.0$, and $\Delta = 0.1$. }   \label{Fig4}
\end{figure}

Besides the thermodynamical averages, the accuracy in the dynamical properties of AIM is also examined. In Fig.4, we show the evolution of LDOS with $U$ at an intermediate temperature $T=0.1$ and $\delta \omega = 0$. The pLacroix results are compared with the NRG results.
For $U$ values ranging from zero to $U=4.0$, we find quantitative agreement between pLacroix and NRG, with the most significant deviation occurring at intermediate $U \approx 2.0$. Compared to NRG, pLacroix produces slightly lower Kondo peak and higher Hubbard peaks. Note that NRG tends to overbroaden the Hubbard peaks. The Kondo peak from pLacroix depends slightly on the value of $\eta$, the broadening of the $\delta$-peaks in the spectral function. Smaller $\eta$ with larger $n_{k}$ and $s$ tends to increase $\rho_{\sigma}(0)$.

\begin{figure}[t!]
\vspace{-1.0cm}
\begin{center}
\includegraphics[width=5.0in, height=3.8in, angle=0]{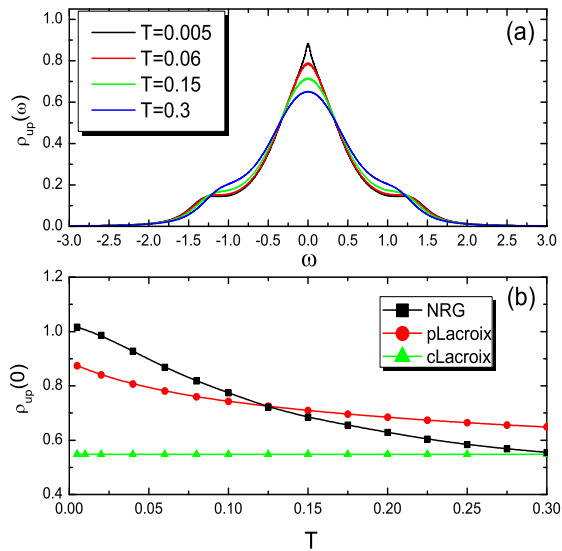}
\vspace*{-2.5cm}
\end{center}
\caption{(a) LDOS obtained from pLacroix at different temperatures. (b) $\rho(0)$ as functions of temperature. Other parameters are $U=1.0$, $\epsilon_d = -U/2.0$, $\delta \omega = 0.0$, and $\Delta =0.1$. The numerical parameters are $s=0.2$ and $\eta=0.01$. }   \label{Fig5}
\end{figure}

The temperature dependence of the LDOS from pLacroixs and NRG is shown in Fig.5. It was proved that cLacroix produces $T$-independent LDOS at the particle-hole symmetric point and paramagnetic bath.\cite{Kashcheyevs1} This severe drawback is improved in pLacroix which produces qualitatively correct evolution of LDOS with temperature, as shown in Fig.5(a) for $U=1.0$. The quantitative comparison of $\rho_{\uparrow}(0)-T$ curve is shown in Fig.5(b). pLacroix produces the correct decreasing function of $\rho(0)(T)$, but with a weaker temperature dependence, crossing NRG curve at $T=0.12$.

\begin{figure}[t!]
\vspace{-1.0cm}
\begin{center}
\includegraphics[width=5.0in, height=3.8in, angle=0]{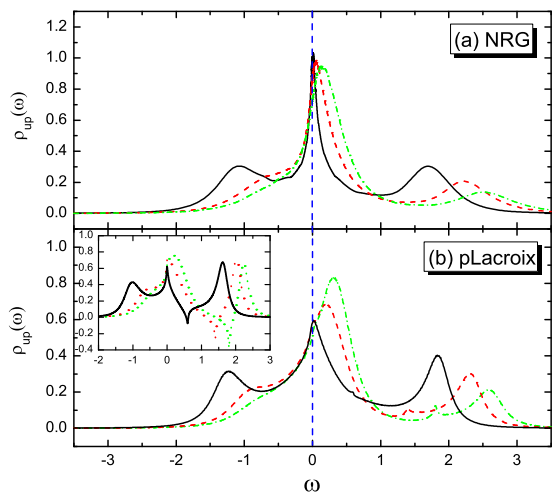}
\vspace*{-3.0cm}
\end{center}
\caption{Impurity density of states from (a) NRG, and (b) pLacroix calculations in the Kondo to mixed valence crossover regime, for the parameters $\epsilon_d=-0.7$ (solid black lines), $\epsilon_d=-0.3$ (red dashed lines), and $\epsilon_d=-0.1$ (green dash-dotted lines). The vertical dashed line marks the Fermi energy. Other parameters are $U=2.0$, $T=0.001$, $\delta \omega = 0.0$, and $\Delta =0.1$. The numerical parameters are $s=0.0$ and $\eta=0.02$. Inset: impurity density of states of cLacroix. }   \label{Fig6}
\end{figure}

Now we switch to the low temperature spectral function. Fig.6 presents a comparison of LDOS at low temperature $T=0.001$, $U=2.0$. For the three values $\epsilon_d = -0.7$, $-0.3$, and $-0.1$, the ground state of AIM crosses over from Kondo regime to mixed valence regime. The corresponding evolution of LDOS is shown in Fig.6(a) for NRG, and in Fig.6(b) for pLacroix. Qualitative consistency is found in the evolution of the central peak and of the Hubbard peaks. As $\epsilon_d$ increases from close to $-U/2$ to zero, the weight of lower Hubbard peak disappears and it transfers to the central peak. The central peak evolves from a Kondo peak at the Fermi energy to a broader quasi-particle peak at a positive frequency. The height of the upper Hubbard peak decreases and its position moves to a higher frequency. Again, compared to NRG, pLacroix produces a lower central peak and sharper Hubbard peaks. The shift of central peak position is more apparent in the pLacroix result. In contrast, in the inset of Fig.6(b), the LDOS from cLacroix has a negative dip at $\omega = 2 \epsilon_d + U$ (Ref.~\onlinecite{Fang1}), which is a signature of violation of causality in the conventional Lacroix approximation.\cite{Note1}

\begin{figure}[t!]
\vspace{-1.0cm}
\begin{center}
\includegraphics[width=5.2in, height=4.0in, angle=0]{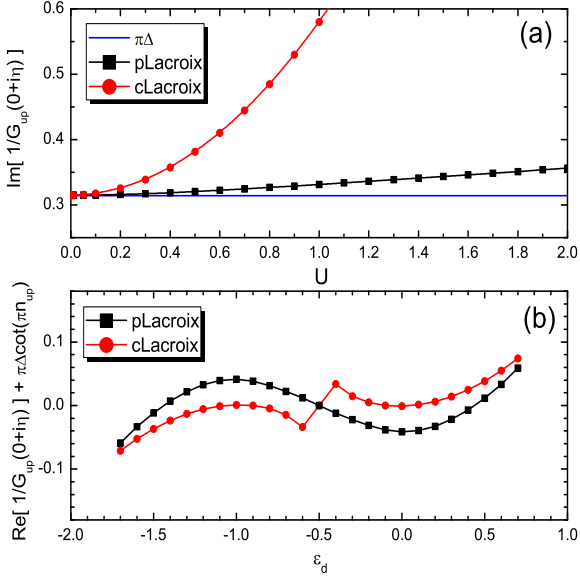}
\vspace*{-2.0cm}
\end{center}
\caption{Checking the Fermi liquid properties in pLacroix and cLacroix results. (a) Checking ${\text Im}G^{-1}_{\uparrow}(0+i\eta) = \pi \Delta$ for different $U$ values with $\epsilon_d=-U/2$. $\eta=10^{-3}$ is used. (b) Checking ${\text Re} [G^{-1}_{\uparrow}(0+i\eta)] = - \pi \Delta {\text cot}(\pi n_{\uparrow} )$ in the wide band limit $\omega_c=200.0$ for different $\epsilon_d$ values at $U=1.0$. Other parameters are $T=0.0$, $\delta \omega = 0.0$, and $\Delta=0.1$.}   \label{Fig7}
\end{figure}

In Fig.7, we compare the accuracy of pLacroix and cLacroix by examining to what extent are the Fermi liquid properties obeyed in the local GF. Two independent Fermi liquid properties are checked\cite{Langreth1} for the paramagnetic bath at $\delta \omega = 0$ and $T=0$. The first one is the unitary condition related to the height of the LDOS. For our hybridization function Eq.(25), it reads\cite{Kashcheyevs1}
\begin{equation}      \label{51}
   {\text Im} [ G_{\sigma}^{-1}(0+i\eta)] = \pi \Delta.
\end{equation}
The second one is the Friedel sum rule. In the wide-band limit, it reads\cite{Kashcheyevs1}
\begin{equation}      \label{52}
{\text Re} [G_{\sigma}^{-1}(0+i\eta)] = - \pi\Delta {\text cot}(\pi \langle n_{\sigma} \rangle  ).
\end{equation}
Fig.7(a) shows that  at particle-hole symmetry, pLacroix result fulfills the first equation much better than cLacroix does in the full $U$ axis. The deviation from $\pi \Delta$ in both curves shows that the Fermi liquid properties of AIM in the Kondo regime is broken at various extents by pLacroix and cLacroix. In Fig.7(b), it is seen that the Friedel sum rule is fulfilled not exactly, with similar errors for the two approximations. The relative errors are less than $10\%$ at $\epsilon_d + U/2 = 1.3$. The discontinuity in cLacroix curve at $\epsilon_d=-U/2$ shows that the particle-hole symmetric point is singular in the conventional Lacroix approximation.\cite{Kashcheyevs1}
To respect the Fermi liquid properties satisfactorily, the truncation approximation should be exact at least up to $U^{2}$ order, as shown by the weak-coupling perturbation theory.\cite{Yamada1,Tong1} In the projective truncation scheme, this requires that all the operators containing two and three single particle operators of bath electrons, such as $d_{\bar{\sigma}}^{\dagger} c_{k \bar{\sigma}} c_{p\sigma}$ and $c_{q \bar{\sigma}}^{\dagger} c_{k \bar{\sigma}} c_{p\sigma}$, are included in the basis. Clearly, pLacroix is not exact at the $U^{2}$ level. This points to a direction of improving the results further in the future.

\subsection{Kondo screening and SU(2) symmetry}

\begin{figure}[t!]
\vspace{-1.0cm}
\begin{center}
\includegraphics[width=4.5in, height=3.5in, angle=0]{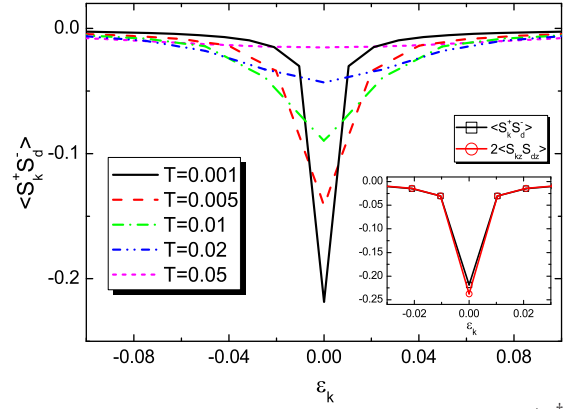}
\vspace*{-3.0cm}
\end{center}
\caption{The spin-spin correlation function $\langle S_{k}^{\dagger} S_{d}^{-}\rangle = \langle c_{k \uparrow}^{\dagger} c_{k \downarrow} d_{\downarrow}^{\dagger} d_{\uparrow} \rangle$ of impurity and bath electrons as functions of $\epsilon_{k}$ for different temperatures. The parameters are $U=2.0$, $\epsilon_d=-U/2$, $\delta \omega = 0.0$, and $\Delta =0.1$. Inset: comparison of $\langle S_{k}^{\dagger} S_{d}^{-}\rangle$ and $2 \langle S_{kz} S_{dz} \rangle = \langle n_{k \uparrow}(n_{\uparrow} - n_{\downarrow}) \rangle$ at $T=0.001$. }   \label{Fig8}
\end{figure}

In the Kondo regime, the impurity spin is screened by forming a spin singlet with the spins of bath electrons. Such nonlocal antiferromagnetic correlations cannot be calculated easily with existing methods such as NRG, hierarchical equation of motion method,\cite{Li1} and other conventional truncation methods for EOM. In the present approach with Lacroix basis, the static spin correlation functions appear in the inner product matrix ${\bf I}$ and hence are naturally obtainable. We also examine the conserving of spin SU(2) symmetry of AIM Hamiltonian at $\delta \omega = 0$. This symmetry guarantees the equivalence between the transverse and longitudinal spin-spin correlation functions. 

In Fig.8, the average of transverse spin exchange $\langle S_{k}^{\dagger} S_{d}^{-} \rangle = - \langle A_{2k \uparrow}^{\dagger} A_{5k \uparrow} \rangle$ is shown as functions of $\epsilon_{k}$ for different temperatures. It is seen that this coupling is anti-ferromagnetic, as expected for the Kondo screening. The energy distribution of the Kondo screening has a peak at the Fermi energy which becomes sharper as temperature decreases, showing the decisive role played by the Fermi surface in the Kondo effect. This calculations can be extended to study the spatial distribution of the Kondo screening in a given lattice geometry, {\it i.e.}, the Kondo cloud. Detailed study of this issue will be the subject of a future work.

Due to the spin SU(2) symmetry of AIM at $\delta \omega=0$, the transverse spin exchange should equal to the longitudinal one,
\begin{equation}      \label{53}
   \langle S_{k}^{\dagger} S_{d}^{-} \rangle = 2 \langle S_{k z} S_{d z}\rangle,
\end{equation} 
where $S_{k}^{\dagger} = c_{k \uparrow}^{\dagger} c_{k \downarrow}$, $S_{d}^{-} = d_{\downarrow}^{\dagger} d_{\uparrow}$, $S_{kz} = (n_{k \uparrow} - n_{k \downarrow})/2$, and $S_{d z} = (n_{\uparrow} - n_{\downarrow} )/2$. 
Since the Lacroix basis Eq.(28) is not spin-rotation invariant, the whole theory of pLacroix does not respect the full SU(2) symmetry, but only the $U(1)$ symmetry associated with the conserving quantity $S^{z}_{tot}$. Therefore, we would expect that the SU(2) symmetry is broken in the pLacroix results. The inset of Fig.8 shows that this breaking of SU(2) symmetry is quite weak. The strongest deviation occurs at the Fermi energy and the relative error is about $10 \%$ at $T=0.001$. It is the task of future work to explore how to recover the exact spin rotation symmetry in the projective truncation approximation.

\section{Discussion and Summary} 

An important issue in the projective truncation is the inner product of two operators. Although in the present work we used the definition Eq.(5), it is clear that any definition is valid, if only it satisfies Eq.(6) and keeps the Liouville matrix ${\bf L}$ Hermitian. The latter requirement is from the time-translation invariance of the given equilibrium state. For an example, in the work of Mori, the inner product was defined as
\begin{equation}      \label{54}
 (A|B) = \frac{1}{\beta} \int_{0}^{\beta} \langle e^{\lambda H} B e^{-\lambda H} A^{\dagger} \rangle d \lambda.
\end{equation}
This definition gives the most probable path of time evolution of operators when the higher order effect is neglected.\cite{Mori1}
Different selection of the inner product amounts to different criterion for the relative importance of basis operators, and directly influences the efficiency of the method. For an example, here we could use Eq.(5) at the infinite temperature limit as an easy-to-calculate inner product. Such a definition, though valid and simple, is not good for the low temperature accuracy of the calculation. The optimal selection of the inner product is thus an important issue for improving the projective truncation method.

The projective truncation scheme introduced in this work is a general method, being applicable not only to AIM, but also to other models of interest, such as Hubbard model and periodic Anderson model. The extension of this theory to models of interacting bosons or spin systems are also possible.\cite{Rowe1} The key of solving these models is the selection of basis operators. Physically, the basis operators should be as close as possible to the annihilation operators of the physical elementary excitations. For the model Hamiltonian defined on a periodic lattice, the spatial translation symmetry can be used to simplify the calculation. In this direction, the present theory could go beyond the two-pole approximation\cite{Roth1} and its extensions \cite{Avella1} for Hubbard model by including more basis operators for a given momentum.

The projective truncation proposed here can be extended to study the quantum quench problems~\cite{Ochoa1} or to the Keldysh GF~\cite{Haug1} without much modification.It is so because this method is a thermodynamics-from-dynamics approach, in which the dynamics of the basis operators are obtained first and the thermodynamical properties are obtained from them. Using the obtained generalized eigenvector and eigen-energy of Liouville matrix, it is easy to construct the time evolution of the basis operators in the Heisenberg picture using, for example, the Laplacian transformation. Compared with the equilibrium problem studied here, in the quantum quench problem, the average of time-dependent operators should be calculated on an initial density matrix. 

Developing a fast and accurate impurity solver for DMFT is one of the motivations of this work. The EOM method has been used for this purpose. Due to the deficiencies discussed above, conventional Lacroix approximation may have problems as an impurity solver.\cite{Zhu1} Luo carried out EOM decoupling approximation beyond Lacroix level using the formalism of conected GFs, \cite{Luo1} giving improved results. However, this approach is quite complicated and is difficult to implement for the multi-orbital case. Other attempts in this direction received only partial success.\cite{Feng1,Tong2} With the projective truncation method proposed here, we can think of developing a well-behaved impurity solver for general multi-orbital AIMs and find applications in DMFT study of lattice models.

In summary, in this paper we present a practical method to implement the operator projection theory for the EOM of GFs. This scheme does not have the arbitrariness in the Tyablikov-type decoupling approximation. The frequently encountered problem of causality violation is avoided from the outset. Compared to previous theories based on operator truncation idea, we reduce the problem of evaluating the Liouville matrix to the calculation of inner product matrix ${\bf I}$ and the natural closure matrix ${\bf M}$, by introducing the partial projection approximation for ${\bf L}$. In this process, the correlations are kept as much as possible by employing the exact identities from time translation invariance. We applied this method to AIM on the Lacroix basis. Comparison to results from conventional Lacroix approximation and NRG shows that the partial projection truncation improves over the conventional Lacroix approximation. The issue of inner product and the applicability of present method to other quantum many-body problems are discussed.

\section{Acknowledgements}
This work is supported by 973 Program of China (2012CB921704), NSFC grant (11374362), Fundamental Research Funds for the Central Universities, and the Research Funds of Renmin University of China 15XNLQ03. NHT acknowledge helpful discussions with T. F. Fang and Y. Qi.

\appendix{}

\section{Identities of Averages from Hermiticity of ${\bf L}$ }

In this Appendix, we summarize the exact identities resulting from the Hermiticity of ${\bf L}$ matrix. From definition ${\bf L}_{ij} = (A_{i}|[A_{j}, H])$, the Hermitian condition ${\bf L}_{ij} = {\bf L}_{ji}^{\ast}$ can be shown to be equivalent to 
\begin{equation}      \label{A1}
 \langle [ \{A_{i}^{\dagger}, A_{j} \}, H ]\rangle = 0.
\end{equation}
Using $J_{ij} \equiv \{ A_{i}^{\dagger}, A_{j} \}$ to represent the conserving operators, below, we summarize the non-trivial identities $\langle [J_{ij}, H]\rangle = 0$ for the Lacroix basis Eq.(28). The fact that the averages are real was used to simplify the equations. First, we find that $J_{5k, 1} = c_{k \bar{\sigma}}^{\dagger} d_{\bar{\sigma}}$ and $J_{6k, 1} = d_{\bar{\sigma}}^{\dagger} c_{k \bar{\sigma}} $ give the equivalent non-trivial identity, 
\begin{eqnarray}      \label{A2}
&& (\epsilon_d - \epsilon_{k \bar{\sigma}} ) \langle  c_{k \bar{\sigma}}^{\dagger} d_{\bar{\sigma}} \rangle  - V_{k \bar{\sigma}} \langle n_{\bar{\sigma}} \rangle + U \langle n_{\sigma} c_{k \bar{\sigma}}^{\dagger} d_{\bar{\sigma} } \rangle  \nonumber \\
&& + \sum_{p} V_{p \bar{\sigma}} \langle c_{k \bar{\sigma}}^{\dagger}c_{p \bar{\sigma} } \rangle = 0.
\end{eqnarray}
This equation has the meaning of electron current conservation in the equilibrium condition. 

Another non-trivial identity comes from $J_{5k, 3} = n_{\sigma} c_{k \bar{\sigma}}^{\dagger}d_{\bar{\sigma}}$ and it means the conservation of impurity-change-correlated electron current,
\begin{eqnarray}      \label{A3}
&& (\epsilon_d - \epsilon_{k \bar{\sigma}} + U) \langle n_{\sigma} c_{k \bar{\sigma}}^{\dagger} d_{\bar{\sigma}} \rangle  - V_{k \bar{\sigma}} \langle n_{\bar{\sigma}} n_{\sigma} \rangle   \nonumber \\
&& + \sum_{p} V_{p \bar{\sigma}} \langle n_{\sigma} c_{k \bar{\sigma}}^{\dagger} c_{p \bar{\sigma}}  \rangle  \nonumber \\
&& + \sum_{p} V_{p \sigma} \langle ( d_{\sigma}^{\dagger}c_{p\sigma} - c_{p\sigma}^{\dagger} d_{\sigma} ) c_{k \bar{\sigma}}^{\dagger }d_{\bar{\sigma}} \rangle  =0 .
\end{eqnarray}
From $J_{6k,3} = (1-n_{\sigma}) d_{\bar{\sigma}}^{\dagger} c_{k \bar{\sigma}}$, one can obtain the particle-hole symmetric correspondence of Eq.(A3) as,
\begin{eqnarray}      \label{A4}
&& (\epsilon_d - \epsilon_{k \bar{\sigma}} ) \langle (1-n_{\sigma}) d_{ \bar{\sigma} }^{\dagger} c_{k \bar{\sigma} }\rangle  - V_{k \bar{\sigma} } \langle (1- n_{\sigma}) n_{\bar{\sigma} } \rangle   \nonumber \\
&& + \sum_{p} V_{p \bar{\sigma} } \langle (1- n_{ \sigma }) c_{p \bar{\sigma}}^{\dagger} c_{k \bar{\sigma} }  \rangle  \nonumber \\
&& + \sum_{p} V_{p \sigma } \langle (d_{ \sigma }^{\dagger} c_{p \sigma } - c_{p \sigma}^{\dagger} d_{ \sigma } ) d_{\bar{ \sigma} }^{\dagger} c_{k \bar{\sigma} } \rangle  =0 .   
\end{eqnarray}

From $J_{5k, 4p} = -  d_{\sigma}^{\dagger} d_{\bar{\sigma}} c_{k \bar{\sigma}}^{\dagger} c_{p \sigma}$, one obtains the spin-current conservation identity as
\begin{eqnarray}      \label{A5}
&& (\epsilon_{k \bar{\sigma}} - \epsilon_{p \sigma}) \langle d_{\sigma}^{\dagger} d_{\bar{\sigma}} c_{k \bar{\sigma}}^{\dagger} c_{p \sigma} \rangle  + V_{p\sigma} \langle n_{\sigma} c_{k \bar{\sigma}}^{\dagger} d_{\bar{\sigma}} \rangle  \nonumber \\
&& + V_{k \bar{\sigma}} \langle (1- n_{\bar{\sigma}}) d_{\sigma}^{\dagger} c_{p \sigma} \rangle + \sum_{q} V_{q \sigma } \langle c_{q \sigma}^{\dagger} d_{\bar{\sigma}}  c_{k \bar{\sigma}}^{\dagger} c_{p \sigma}  \rangle  \nonumber \\
&& - \sum_{q} V_{q \bar{\sigma}} \langle d_{\sigma}^{\dagger} c_{q \bar{\sigma}} c_{k \bar{\sigma}}^{\dagger} c_{p \sigma}\rangle = 0. 
\end{eqnarray}

From $J_{6k, 4p} = d_{ \bar{\sigma}}^{\dagger} c_{k \bar{\sigma}} c_{p \sigma} d_{\sigma}^{\dagger} $, one obtains the conservation of the pair-hopping current,
\begin{eqnarray}      \label{A6}
&& (\epsilon_{p \sigma} + \epsilon_{k \bar{\sigma}} - 2 \epsilon_d - U) \langle d_{\bar{\sigma}}^{\dagger} d_{\sigma}^{\dagger}  c_{k \bar{\sigma}} c_{p \sigma} \rangle  - V_{p \sigma} \langle n_{\sigma} d_{\bar{\sigma}}^{\dagger} c_{k \bar{\sigma}} \rangle  \nonumber \\
&& - V_{k \bar{\sigma}} \langle n_{\bar{\sigma}} d_{\sigma}^{\dagger} c_{p \sigma} \rangle  - \sum_{q} V_{q \sigma} \langle d_{\bar{\sigma}}^{\dagger} c_{q \sigma}^{\dagger}  c_{k \bar{\sigma}} c_{p \sigma}  \rangle  \nonumber \\
&& - \sum_{q} V_{q \bar{\sigma}} \langle c_{q\bar{\sigma}}^{\dagger}d_{\sigma}^{\dagger} c_{k \bar{\sigma}}  c_{p \sigma} \rangle  = 0. 
\end{eqnarray}
From $J_{5k,5p} = (1- n_{\sigma})n_{\bar{\sigma}} \delta_{kp} + (n_{\sigma}- n_{\bar{\sigma}} ) c_{k \bar{\sigma}}^{\dagger} c_{p \bar{\sigma}} $, one obtains the conservation of the impurity-spin-correlated electron current,
\begin{eqnarray}      \label{A7}
&& (\epsilon_{p \bar{\sigma}} - \epsilon_{k \bar{\sigma}}) \langle ( n_{\sigma}- n_{\bar{\sigma}} ) c_{k \bar{\sigma}}^{\dagger} c_{p \bar{\sigma}}\rangle + V_{p \bar{\sigma}}  \langle n_{\sigma} c_{k \bar{\sigma}}^{\dagger} d_{\bar{\sigma}} \rangle  \nonumber \\
 && +   V_{k \bar{\sigma}}  \langle (1- n_{\sigma} ) d_{\bar{\sigma}}^{\dagger} c_{p \bar{\sigma}} \rangle  + \sum_{q} V_{q \sigma} \langle ( d_{\sigma}^{\dagger} c_{q \sigma} - c_{q \sigma}^{\dagger} d_{\sigma} ) c_{k \bar{\sigma}}^{\dagger} c_{p \bar{\sigma}} \rangle   \nonumber \\
&& -  \sum_{q} V_{q \bar{\sigma}} \langle ( d_{\bar{\sigma}}^{\dagger} c_{q \bar{\sigma}} - c_{q \bar{\sigma}}^{\dagger} d_{ \bar{\sigma}} ) c_{k \bar{\sigma}}^{\dagger} c_{p \bar{\sigma}} \rangle   = 0. 
\end{eqnarray}
From $J_{6k,6p} = n_{\sigma}n_{\bar{\sigma}} \delta_{kp} + (1 - n_{\sigma}- n_{\bar{\sigma}} ) c_{p \bar{\sigma}}^{\dagger} c_{k \bar{\sigma}} $, one obtains the conservation of the impurity-charge-correlated electron current,
\begin{eqnarray}      \label{A8}
&& (\epsilon_{k \bar{\sigma}} - \epsilon_{p \bar{\sigma}}) \langle ( 1- n_{\sigma}- n_{\bar{\sigma}} ) c_{p \bar{\sigma}}^{\dagger} c_{k \bar{\sigma}}\rangle + V_{k \bar{\sigma}}  \langle (1-n_{\sigma}) c_{p \bar{\sigma}}^{\dagger} d_{\bar{\sigma}} \rangle  \nonumber \\
 && + V_{p \bar{\sigma}} \langle n_{\sigma} d_{\bar{\sigma}}^{\dagger} c_{k \bar{\sigma}} \rangle  - \sum_{q} V_{q \sigma} \langle ( d_{\sigma}^{\dagger} c_{q \sigma} - c_{q \sigma}^{\dagger} d_{\sigma} ) c_{p \bar{\sigma}}^{\dagger} c_{k \bar{\sigma}} \rangle   \nonumber \\
&& + \sum_{q} V_{q \bar{\sigma}} \langle ( d_{\bar{\sigma}}^{\dagger} c_{q \bar{\sigma}} - c_{q \bar{\sigma}}^{\dagger} d_{ \bar{\sigma}} ) c_{p \bar{\sigma}}^{\dagger} c_{k \bar{\sigma}} \rangle   = 0. 
\end{eqnarray}
The above seven identities are the exact relations that we implicitly employed in the projection approximation of ${\bf L}$ for the Lacroix basis.

\section{Commutators of $[A_{i}, H]$ }

In this Appendix, we summarize the commutators between the basis operators and AIM Hamiltonian $H$ Eq.(24). For the Lacroix basis, we need the commutators between $A_{i}$ and $H_{e/o}$ of Eq.(40). The commutators with $H_{e}$ are
\begin{eqnarray}      \label{B1-B6}
\left[ d_{\sigma}, H_{e} \right] &=& -\frac{U}{2} d_{\sigma} + \sum_{k} V_{k\sigma} c_{k\sigma}  + U n_{\bar{\sigma}} d_{\sigma} ,  \\
\left[ c_{k \sigma}, H_{e} \right] &=& \epsilon_{k \sigma} c_{k \sigma} + V_{k \sigma} d_{\sigma},   \\
\left[ n_{\bar{\sigma}}d_{\sigma}, H_{e} \right] &=& \frac{U}{2} n_{\bar{\sigma}}d_{\sigma} + \sum_{k} V_{k\sigma} n_{\bar{\sigma}} c_{k\sigma}   \nonumber\\
&& + \sum_{k} V_{k \bar{\sigma}} ( d_{\bar{\sigma}}^{\dag} c_{k\bar{\sigma}} - c_{k\bar{\sigma}}^{\dag}d_{\bar{\sigma}} ) d_{\sigma},   \\
\left[ n_{\bar{\sigma}} c_{k\sigma}, H_{e} \right] &=&  \epsilon_{k\sigma} n_{\bar{\sigma}} c_{k \sigma} + V_{k \sigma} n_{\bar{\sigma}} d_{\sigma}\nonumber\\
&& + \sum_{p} V_{p \bar{\sigma}} ( d_{\bar{\sigma}}^{\dag} c_{p \bar{\sigma}}  - c_{p \bar{\sigma}}^{\dag} d_{\bar{\sigma}} ) c_{k\sigma} ,  \\
\left[ d_{\bar{\sigma}}^{\dag}  c_{k \bar{\sigma}}d_{\sigma}, H_{e} \right] &=& 
\epsilon_{k \bar{\sigma}} d_{\bar{\sigma}}^{\dag}  c_{k \bar{\sigma}}d_{\sigma}  + V_{k \bar{\sigma}} n_{\bar{\sigma}}d_{\sigma} - \frac{1}{2}V_{k \bar{\sigma}} d_{\sigma}  \nonumber \\
&& + \sum_{p }V_{p \sigma} d_{\bar{\sigma}}^{\dagger} c_{k \bar{\sigma}} c_{p \sigma}  \nonumber \\
&& - \sum_{p }V_{p \bar{\sigma}} ( c_{p \bar{\sigma}}^{\dag} c_{k \bar{\sigma}} - \frac{1}{2} \delta_{kp} ) d_{\sigma},  \\
\left[c_{k \bar{\sigma}}^{\dag} d_{\bar{\sigma}}d_{\sigma}, H_{e} \right] &=& 
-\epsilon_{k \bar{\sigma}} c_{k \bar{\sigma}}^{\dag} d_{\bar{\sigma}}d_{\sigma}  - V_{k \bar{\sigma}} n_{\bar{\sigma}}d_{\sigma} + \frac{1}{2}V_{k \bar{\sigma}} d_{\sigma}  \nonumber \\
&& + \sum_{p }V_{p \sigma} c_{k \bar{\sigma}}^{\dag} d_{\bar{\sigma}} c_{p \sigma}  \nonumber \\
&& + \sum_{p }V_{p \bar{\sigma}} ( c_{k \bar{\sigma}}^{\dag} c_{p \bar{\sigma}} - \frac{1}{2} \delta_{kp} ) d_{\sigma}. 
\end{eqnarray}
The commutators with $H_{o}$ are
\begin{eqnarray}      \label{B7-B12}
\left[d_{\sigma}, H_{o} \right] &=& \left(\epsilon_{d} -\mu + \frac{U}{2} \right) d_{\sigma}, \\
\left[c_{k \sigma}, H_{o} \right] &=& -\mu c_{k\sigma}, \\
\left[n_{\bar{\sigma}}d_{\sigma}, H_{o} \right] &=&
\left(\epsilon_{d} - \mu +\frac{U}{2} \right) n_{\bar{\sigma}}d_{\sigma},  \\
\left[ n_{\bar{\sigma}} c_{k \sigma}, H_{o} \right] &=& -\mu n_{\bar{\sigma}}c_{k \sigma}, \\
\left[ d_{\bar{\sigma}}^{\dag} c_{k \bar{\sigma}} d_{\sigma}, H_{o} \right] &=& -\mu d_{\bar{\sigma}}^{\dag}  c_{k \bar{\sigma}} d_{\sigma}, \\
\left[ c_{k \bar{\sigma}}^{\dag} d_{\bar{\sigma}} d_{\sigma}, H_{o} \right] &=& \left( 2 \epsilon_{d} -\mu + U \right) c_{k \bar{\sigma}}^{\dag} d_{\bar{\sigma}} d_{\sigma}.
\end{eqnarray}

\section{Proof of Particle-Hole Symmetry in the Projective Truncation}

In this Appendix, we prove that for a particle-hole symmetric Hamiltonian $H^{\prime}=H$, the GFs obtained by the projective truncation approximation Eq.(10) fulfil the particle-hole symmetry, if Eqs.(36) and (37) are satisfied. At particle-hole symmetric point, $H_{e}=H$ and $H_o = 0$. Eqs.(36) and (37) becomes the symmetry condition for ${\bf I}$, ${\bf M}$, ${\bf P}$, and ${\bf L}$. The particle-hole transformation is defined in Eq.(33). First, we give the relation that a particle-hole symmetric GF must obey, and then we show that Eq.(10) indeed produces GFs fulfilling this relation.

For a particle-hole symmetric Hamiltonian $H$, it is easy to prove that the average of an operator and a GF have the following properties,
\begin{eqnarray}      \label{C1}
&& \langle O^{\prime} \rangle = \langle O \rangle,  \nonumber \\
&& G(A|B)_{\omega} = G( A^{\prime}|B^{\prime} )_{\omega}.
\end{eqnarray}
From Lehmann representation of GF, we also get the relation
\begin{equation}      \label{C2}
   G (A^{\dagger}|B^{\dagger} )_{\omega} = - G^{\ast}(A|B)_{-\omega}.
\end{equation}
Note that the complex conjugate only applies to the matrix element in the GF, not to the frequency $\omega + i \eta$ in the retarded GF. 

Combining Eqs.(C1) and (C2) and applying it to the operators $\bar{A}$ and $\bar{B}$, we obtain 
\begin{equation}      \label{C3}
   G(\bar{A}| \bar{B} )_{\omega} = - G^{\ast}(\tilde{A} | \tilde{B} )_{-\omega}.
\end{equation}
Here we have used the definition Eq.(34) for $\tilde{O}$. $\bar{O}$ is defined below Eq.(34).
We define the GF matrix as $[{\bf G}(\omega)]_{ij} = G(A_i | A_{j}^{\dagger} )_{\omega}$ and 
$\left[ \bar{ {\bf G} } (\omega) \right]_{ij} = G( \bar{A}_i | \bar{A}_{j}^{\dagger} )_{\omega}$. 
Using $\vec{ \tilde{A} } = {\bf Q} \vec{A}$, we obtain from Eq.(C3) that
\begin{equation}      \label{C4}
   \bar{ {\bf G} }(\omega) = - {\bf Q}^{\ast} {\bf G}^{\ast}( - \omega) {\bf Q}^{T}.
\end{equation}
This is the particle-hole symmetry properties of GF. In particular, using the ${\bf Q}$ matrix in Eq.(35), we have
\begin{equation}      \label{C5}
   G(d_{\bar{\sigma}} | d_{\bar {\sigma}}^{\dagger} )_{\omega} = - G(d_{\sigma} | d_{\sigma}^{\dagger} )_{-\omega},
\end{equation}
which leads to the relation for LDOS $\rho_{\bar{\sigma}}(\omega) = \rho_{\sigma}(-\omega)$.

Below we prove that Eq.(10) produces GFs that satisfy this symmetry condition.
Writing down the EOM for the GF $G(\bar{A}_{i} | \bar{A}_{j} )_{\omega}$ and using the definitions 
$\bar{ {\bf I}}_{ij} = ( \bar{A}_{i} | \bar{A}_{j} )$, $\bar{ {\bf L}}_{ij} = ( \bar{A}_{i} | [ \bar{A}_{j}, H ] )$, 
and $\xoverline[0.7]{{\bf M}}_{t} = \bar{ {\bf I}}^{-1} \bar{ {\bf L}}$, we obtain the correspondence of Eq.(10) for $\bar{ {\bf G} }(\omega)$ as
\begin{equation}      \label{C6}
 \bar{ {\bf G}} (\omega) \approx \left( \omega {\bf 1} - \xoverline[0.7]{ {\bf M} }_{t}^{T} \right)^{-1} \bar{ {\bf I}}^{T}.
\end{equation}
At the particle-hole symmetric point, the matrices $\bar{ {\bf I} }$ and $\bar{ {\bf L} }$ should fulfil the following relations,
\begin{eqnarray}      \label{C7}
  \bar{{\bf I}}_{ij} &\equiv& ( \bar{A}_{i} | \bar{A}_{j} )    \nonumber \\
  &=& ( \bar{A}^{\prime}_{i} | \bar{A}^{\prime}_{j} ) = ( \tilde{A}_{i} | \tilde{A}_{j} )^{\ast} = (\tilde{{\bf I}})^{\ast}_{ij},
\end{eqnarray}
and
\begin{eqnarray}      \label{C8}
  \bar{{\bf L}}_{ij} &\equiv& ( \bar{A}_{i} | [\bar{A}_{j}, H] )   \nonumber \\
  & =& ( \bar{A}^{\prime}_{i} | [\bar{A}^{\prime}_{j}, H] ) = -( \tilde{A}_{i} | [\tilde{A}_{j}, \tilde{H}] )^{\ast} = (\tilde{{\bf L}})^{\ast}_{ij}.
\end{eqnarray}
From them, we obtain $\xoverline[0.7]{{\bf M}}_{t} = \tilde{{\bf M}}_{t}^{\ast}$. Putting these relations into Eq.(C6) and using Eqs.(36) and (37) which hold for particle-hole symmetric $H$, it is easy to obtain the particle-hole symmetry properties of GFs Eq.(C4). Since in our method, the inner product matrix ${\bf I}$ and the Liouville matrix ${\bf L}$ are obtained self-consistently from the GFs, a particle-hole symmetric GF matrix will guarantee the validity of Eqs.(C7) and (C8). As a result of self-consistency, our projective truncation method will conserve the particle-hole symmetry in the GF, provided that ${\bf I}$ and ${\bf L}$ are calculated exactly from the GFs without further particle-hole symmetry breaking approximations.

\end{document}